\documentclass[12pt]{iopart}

\usepackage{graphicx,color}
\expandafter\let\csname equation*\endcsname\relax
\expandafter\let\csname endequation*\endcsname\relax
\usepackage{amsmath}
\usepackage{amssymb}
\usepackage{hyperref}
\usepackage[normalem]{ulem}

\newcommand{\Up}{{\uparrow}}
\newcommand{\Dn}{{\downarrow}}

\newcommand{\kF}{k_\mathrm{F}}
\newcommand{\pc}{p_\mathrm{C}}
\newcommand{\ps}{p_\mathrm{S}}
\newcommand{\thetas}{\theta_\mathrm{S}}
\newcommand{\thetac}{\theta_\mathrm{C}}
\newcommand{\phis}{\varphi_\mathrm{S}}

\newcommand{\Ks}{K_\mathrm{S}}
\newcommand{\Kc}{K_\mathrm{C}}
\newcommand{\LR}{\mathrm{L,R}}
\newcommand{\RL}{\mathrm{R,L}}

 \newcommand\bwt         {\begin{widetext}}
 \newcommand\ewt         {\end{widetext}}

\newcommand{\dd}{\mathrm{d}} 

\begin{document}

\title{The resonant state at filling factor $\nu=1/2$ in chiral fermionic ladders}

\author{Andreas Haller$^1$, Matteo Rizzi$^1$ and Michele Burrello$^2$}

\address{$^1$Johannes-Gutenberg-Universit\"at Mainz, Institut f\"ur Physik,
Staudingerweg 7, D-55099 Mainz, Germany}

\address{$^2$Niels Bohr International Academy and Center for Quantum Devices, Niels Bohr Institute, University of Copenhagen, Juliane Maries Vej 30,
2100 Copenhagen, Denmark}

\eads{\mailto{hallera@uni-mainz.de}, \mailto{rizzi@uni-mainz.de}, \mailto{michele.burrello@nbi.ku.dk}}

\begin{abstract}
Helical liquids have been experimentally realized in both nanowires and ultracold atomic chains as the result of strong spin-orbit interactions. In both cases the inner degrees of freedom can be considered as an additional space dimension, providing an interpretation of these systems as chiral synthetic ladders, with artificial magnetic fluxes determined by the spin-orbit terms.
In this work, we characterize the helical state which appears at filling $\nu=1/2$:
this state is generated by a gap arising in the spin sector of the corresponding Luttinger liquid and
it can be interpreted as the one-dimensional (1D) limit of a fractional quantum Hall state of bosonic pairs of fermions.
We study its main features, focusing on entanglement properties and correlation functions.
The techniques developed here provide a key example for the study of similar quasi-1D systems
beyond the semiclassical approximation commonly adopted in the description of the Laughlin-like states.
\end{abstract}

\maketitle

\section{Introduction}
The scientific paradigm of topological phases of matter lays its foundation on the experimental engineering of solid state devices ranging from topological insulators and superconductors~\cite{bernevigbook} to fractional quantum Hall setups~\cite{jainbook}. In the last years, however, the striding evolution of this field is progressively investing a variegated plethora of other platforms and, among them, ultracold atoms trapped in optical lattices~\cite{lewensteinbook} offered an unprecedented scenario for the direct implementations of toy-models, such as the Hofstadter~\cite{Aidelsburger2013,Miyake2013,bloch15} or Haldane~\cite{esslinger14} models, which play a key role in our understanding of the topological phenomena in condensed matter physics.

One of the most appealing developments in this field is based on the idea of synthetic dimensions~\cite{Boada2012}: the inner degrees of freedom of the trapped atoms can represent an additional physical dimension and, in this scenario, the introduction of a laser-induced spin-orbit coupling is translated into large magnetic fluxes in the synthetic lattice~\cite{Celi2014}.
This idea has been exploited to create synthetic ladders of fermions~\cite{fallani2015,fallani2016} and bosons~\cite{spielman2015} which, due to the  artificial magnetic fluxes, display features consistent with the presence of gapless helical modes which are interpreted as the 1D limit of the chiral edge modes of a two-dimensional integer quantum Hall state. These gapless ladders can thus be considered an additional tool to investigate the quantum Hall regime, complementing the well-known thin-torus limit of gapped 2D states \cite{bergholtz06}.

The possibility of introducing and tuning interactions in such ultracold atom systems opens the way to the study of the 1D counterpart of the most common fractional quantum Hall (FQH) states. Based on the theory developed for the engineering of FQH states in nanowire arrays~\cite{kane02,teokane}, it was showed that some of the gapless modes of these synthetic ladders can indeed be gapped when the ratio between the artificial magnetic flux per plaquette and the Fermi momentum of the system approaches simple resonant values~\cite{barbarino15,zhai15,lehur15,taddia2016,mazza2016,lehur16,zoller2017}.
These resonant values correspond to the filling fractions $\nu$ of the most common quantum Hall states: a semiclassical analysis based on bosonization~\cite{kane02,teokane,oreg2014} reveals, for example, that Laughlin-like states appear at the expected values $\nu=1/2$ and $\nu=1/3$ for bosons and fermions respectively~\cite{mazza2016}. On the ladder geometry, the main observable to witness the appearance of such states is the chiral current~\cite{sela15}: as a function of either the magnetic flux or the Fermi momentum, the chiral current displays two typical cusps which are spaced out by a linear regime crossing zero exactly at the resonance. These cusps testify the commensurate-incommensurate transitions which determine the rise of the Laughlin-like 1D states~\cite{mazza2016}.

Besides these well-understood cases, the study of the topological adiabatic pumping in synthetic fermionic cylinders~\cite{taddia2016} suggests that other resonant states appear at different filling fractions. The main example emerges at $\nu=1/2$. Such state is qualitatively different from the Laughlin-like states: despite presenting some similarities in its observables (as in the case of the chiral current, see Figs.~\ref{fig:general}-\ref{fig:discr}), it cannot be trivially explained in terms of a semiclassical approximation and its characterization is, so far, unknown.

In this paper we examine in detail this resonant state at filling $\nu=1/2$ in a fermionic ladder. The analysis of its observables on one side, and the renormalization group study of its field theoretical description on the other, suggest that this state is the 1D limit of a strongly paired state: in this regime fermions bind pairwise into effective bosons which arrange themselves in a bosonic Laughlin state at filling $\nu'=1/8$, since there are half as many particles each carrying twice the charge~\cite{greiter91}. This state is often referred to as the $K=8$ state~\cite{wen08} and corresponds to a strong-paired phase in spinless p-wave superconductors~\cite{readgreen}.

\section{The model}
Our analysis is based on the tight-binding description of a two-component fermionic chain, characterized by a spin-orbit coupling with amplitude $\phi$, which can be interpreted as a magnetic flux in the synthetic-dimension picture. The single-particle Hamiltonian reads:
\begin{equation} \label{Hamspace}
 H_{\mathrm{sp}}=  \sum_r{ - t  \left( a^\dag_r \, e^{i\frac{\phi\sigma_z}{2}}\,  a_{r+1} + {\rm H.c.}\right)  + \Omega \, a^\dag_r \, \sigma_x  \, a_r},
\end{equation}
where $a,a^\dag$ are two-component fermionic spinors and $\Omega$ is the rung tunneling,
typically induced through optical tools~\cite{fallani2015,fallani2016}.
We consider $N$ fermions in a chain of length $L$, such that, in the limit $\Omega \to 0$, the Fermi momentum is set to $\kF = \pi \rho_0/2$ with average density $\rho_0=N/L$.
\begin{figure}[ht]
	\centering
	\includegraphics{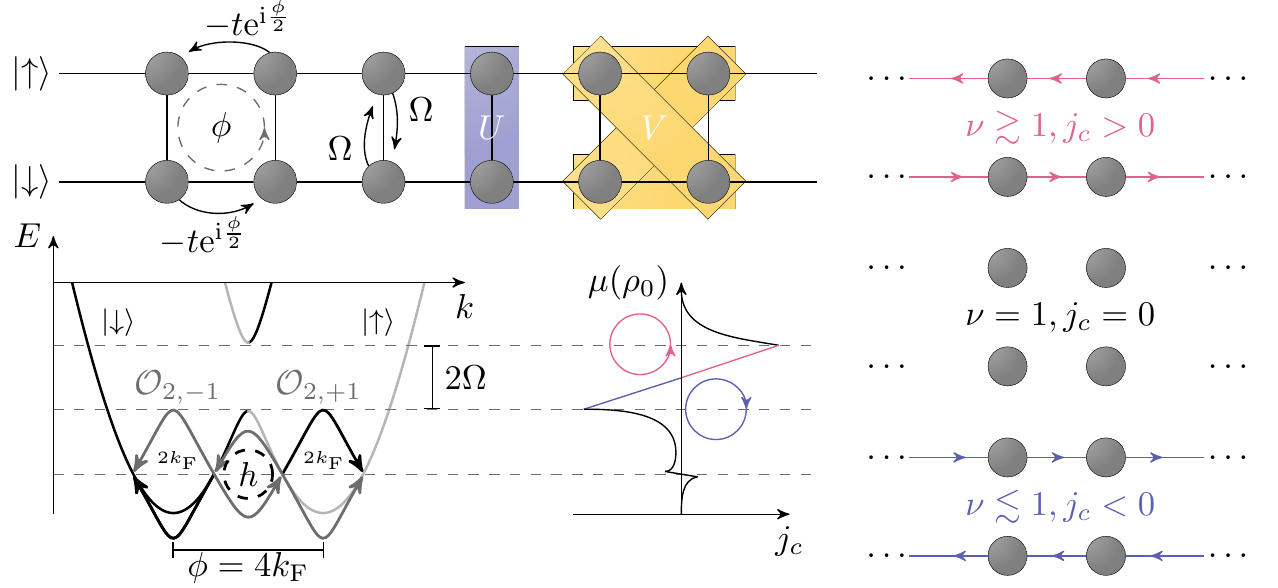}\llap{
  \parbox[b]{12cm}{(a)\\\rule{0ex}{4.4cm}
  }}\llap{
  \parbox[b]{12.25cm}{(b)\\\rule{0ex}{0.5cm}
  }}\llap{
  \parbox[b]{7.8cm}{(c)\\\rule{0ex}{0.5cm}
  }}\llap{
  \parbox[b]{0cm}{(d)\\\rule{0ex}{4.4cm}
  }}\llap{
  \parbox[b]{0cm}{(e)\\\rule{0ex}{2.5cm}
  }}\llap{
  \parbox[b]{0cm}{(f)\\\rule{0ex}{0.6cm}
  }}
\caption{
\label{fig:general}
(a): model Hamiltonian, arrows sketch hopping terms and colored boxes impicture interactions.
(b): dispersion relation of the non-nteracting Hamiltonian.
For generic densities $\rho_0$, there are four linearly dispersing gapless modes,
which correspond to the bosonized fields.
The transverse hopping $\Omega$ opens a (non-interacting) gap for two of them
around the integer resonance $\nu=1$.
(c): chiral current with typical cusp signature and linear regime inside the $\nu=1$ gap.
(d) to (f): sketches of the current flowing in the synthetic dimension inside the non-interacting gap.
For lower densities, around  $\nu=1/2$, the interaction terms $\mathcal{O}^{\LR}_{2,\pm1}$ of Eq.~\eqref{flip1}
can gap out the two internal modes by making the term $h \cos\left(2\sqrt{2}\thetas \right)$ in Eq.~\eqref{Ham12}
relevant, as discussed in the text.
This is reflected in a resonant feature in the chiral current, as shown in detail in Fig.~\ref{fig:discr}.
}
\end{figure}

We model the repulsion among the atoms with a combination of on-site and nearest-neighbor terms:
\begin{equation} \label{HUV}
H_{UV} = \sum_r U n_r (n_r-1)/2 + Vn_rn_{r+1}
\end{equation}
where $n_r= a^\dag_{r,\Up}a_{r,\Up} + a^\dag_{r,\Dn}a_{r,\Dn}$ is the local density, and this interaction is invariant under spin-rotations, as expected in the experiments with $^{173}$Yb atoms, where the two spin species represent different hyperfine states~\cite{fallani2014}.
We observe, however, that the $SU(2)$ spin symmetry is broken by the $\Omega$ term in Eq.~\eqref{Hamspace}.
We will exploit extensive DMRG calculations \cite{White,White2} based on the MPS ansatz~\cite{usualSchollwoeck,anthology}
to tackle the strongly interacting regime, i.e., $U$ and $V$ comparable with the bandwidth of the system.

We focus on the regime $\Omega \ll |2t\tan(\phi/2)\sin(\phi/2)|$,
such that for low densities $\rho_0$ there are four gapless modes with momenta roughly $\pm\phi/2 \pm \kF$
\cite{kane02,footnote1}.
This allows us to describe the system through a bosonized approach~\cite{giamarchi,nersesyanbook}
in terms of two pairs of dual bosonic fields $\varphi_{a}, \theta_{a}$, with $a= \mathrm{C,S}$ for charge and spin, respectively.
The local density can be approximated by $n_r \approx \rho_0 + \sqrt{2} \, \partial_x \thetac /\pi $
and the chiral current by  $j_c \propto \partial_x \phis$.
The most relevant contributions of the Hamiltonian read:
\begin{align} \label{eq:Hampert}
H = H_0 + g \int \dd x \sum_{\pc>0} \sum_{\ps=-\pc+1}^{\pc-1} \left[\mathcal{O}^{\LR}_{\pc,\ps} + \mathcal{O}^{\mathrm{\RL}}_{\pc,\ps}\right]
\end{align}
where $H_0$ is the Luttinger liquid Hamiltonian for the four gapless modes,
the constraints on $\pc, \ps$ are due to the fermionic nature of the constituents,
and the operators $\mathcal{O}$ appear from the mixing term in Eq.~\eqref{Hamspace} ($g \propto \Omega$): 
\begin{align}
&\mathcal{O}^{\LR}_{\pc,\ps} \propto e^{-i \left[(\phi-2 \pc \kF) x + \sqrt{2} (\phis - \pc \thetac - \ps \thetas) \right]}  + {\rm H.c.} \, ,\label{flip1}
\end{align}
with $\mathcal{O}^{\LR}_{\pc,\ps} = \mathcal{O}^{\mathrm{\RL}}_{-\pc, -\ps}$(see~\ref{RG}).
The scaling dimension of such operators is $D_g = (\Ks^{-1} + \ps^2 \Ks + \pc^2 \Kc)/2$,
in terms of the Luttinger parameters $K_{\mathrm{C,S}}$.
However, they are characterized by a fast oscillating behavior in $x$ and are thus irrelevant,
unless the special resonances $\phi=\pm 2 \pc \kF$ are met.
Such resonances can be related to the quantum Hall states at filling $\nu= 2 \kF / \phi = 1/\pc$,
which indeed measures the ratio between the number of particles $N$ and flux quanta $N_\phi = L\phi/2\pi$
in the ladder~\cite{kane02,teokane,sela15,mazza2016}.
Without loss of generality, we deal henceforth with $\phi>0$ and $\mathcal{O}^{\LR}_{\pc,\ps}$ operators only.

When $\ps=0$ (hence $\pc$ is odd), as in most of the existing literature, the mechanism at the resonance is well understood.
$\mathcal{O}^{\LR}_{\pc,0}$ might become relevant,
and thus pin the combination of fields $\phis - \pc \thetac$ to its semiclassical minima.
This opens a gap between two of the four gapless modes through a commensurate-incommensurate phase transition.
The main effect of this gap can be seen by observing a typical double-cusp pattern of the chiral current $j_c$
for small variations of $\phi$ (or $\kF$) around the resonance~\cite{sela15,mazza2016} (see Fig.~\ref{fig:general}).
No interaction is needed to trigger the relevance of $\mathcal{O}^{\LR}_{1,0}$ at the integer filling $\nu=1$,
which indeed corresponds to a gap opening at $k=0$ in the single particle spectrum due to the $\Omega$ term only.
Instead, next-nearest-neighbor repulsions are needed to reduce the scaling dimension of $\mathcal{O}^{\LR}_{3,0}$
below $2$, thus originating a Laughlin-like $\nu = 1/3$ state~\cite{sela15,mazza2016}.

\begin{figure}[t]
	\centering
	\includegraphics{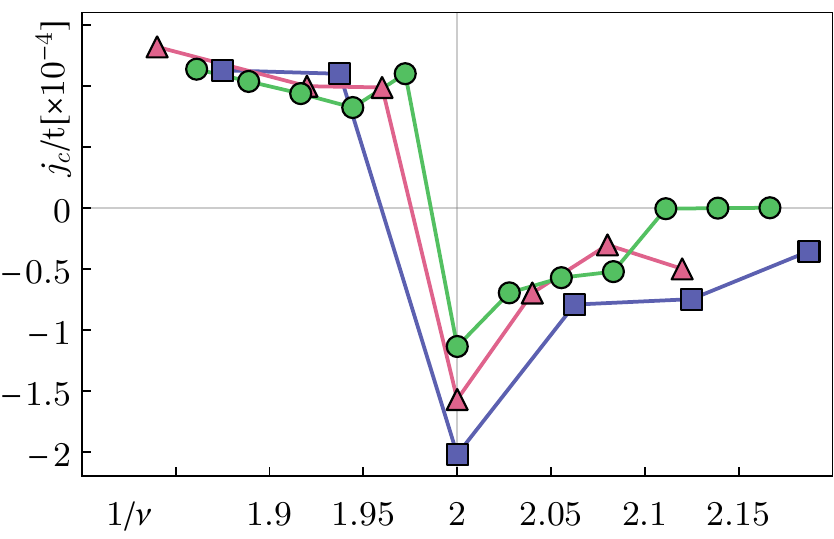}\llap{
  \parbox[b]{2cm}{(a)\\\rule{0ex}{4.4cm}
  }}\\
	\includegraphics{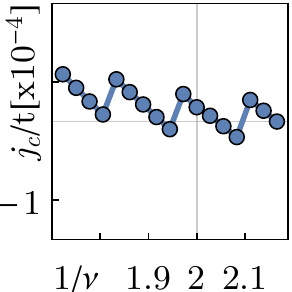}\llap{
  \parbox[b]{1.6cm}{(b)\\\rule{0ex}{.9cm}
  }}\hfill
	\includegraphics{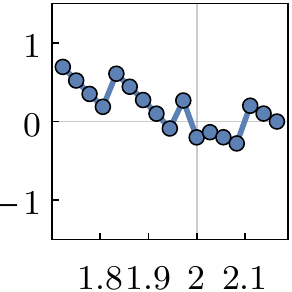}\llap{
  \parbox[b]{1.6cm}{(c)\\\rule{0ex}{.9cm}
  }}\hfill
	\includegraphics{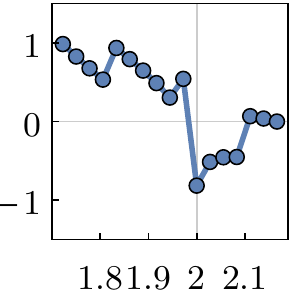}\llap{
  \parbox[b]{1.6cm}{(d)\\\rule{0ex}{.9cm}
  }}\hfill
	\includegraphics{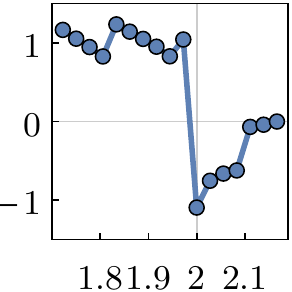}\llap{
  \parbox[b]{1.6cm}{(e)\\\rule{0ex}{.9cm}
  }}\hfill
	\includegraphics{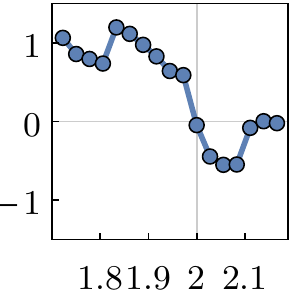}\llap{
  \parbox[b]{1.6cm}{(f)\\\rule{0ex}{.9cm}
  }}
\caption{(a): The chiral current $j_c$ is illustrated in the vicinity of $\nu=1/2$ for system sizes $L/N=72/32$ (square), $110/50$ (triangle) and $156/72$ (disk) at $\Omega/t=0.02$ and $U=V=4t$. The adopted MPS bond dimension is $400$.
The bottom row shows the chiral current for different values of $V$ at $U=4$ at $L/N=156/72$.
(b): $V=1$, (c): $V=2$, (d): $V=3$, (e): $V=4$, (f): $V=5$.
We decided to focus on $U=V=4$ where the sign inversion is the most evident. For discussions of the finite-size scaling at $\nu=1/2$ and related errors see \ref{sec:further_evidence} and \ref{sec:error_discussion}. \label{fig:discr} \label{Fig2}}
\end{figure}

The system is instead more complicated when $\ps\neq 0$ (and $\pc$ is even):
here we focus on the $\nu=1/2$ resonance,
illustrated by the chiral current plot in Fig.~\ref{Fig2}. Similarly to the Laughlin-like states, the chiral current $j_c$ displays a sign inversion across $\nu=1/2$. This is compatible with the appearance of a helical Luttinger state, with counterpropagating gapless modes. The linear behavior of $j_c$ as a function of $\phi-4k_F$ cannot be easily proved, but we qualitatively derive it in \ref{app:d}, see Eq.~(\ref{eq:ohr_current}).
For $\nu=1/2$, indeed, \emph{two} operators $\mathcal{O}^{\LR}_{2,\pm1}$
with the same scaling dimension loose their fast oscillating behavior.
They can be written as $e^{i\Phi_{2,\pm1}}$, with
$\Phi_{2,\pm1}\equiv \sqrt{2} (- \phis + 2 \thetac \pm \thetas)$,
making it apparent that $\Phi_{2,\pm1}$ do not commute with each other;
this hinders the previous semiclassical approximation.
Therefore, we performed the analysis of the renormalization group (RG) flow
at second order in the coupling $g$ (see~\ref{RGFlow}).
Such calculation follows the RG techniques developed for the study of pair-hopping terms in electronic ladders%
~\cite{yakovenko92,nersesyan93,nersesyanbook}.
The main result is the emergence of a new term proportional to
$e^{i\left(\Phi_{2,+1} - \Phi_{2,-1}\right)} + {\rm H.c.} = \cos\left(2\sqrt{2}\thetas \right)$,
which acts in the spin sector only and corresponds to a double backscattering of a pair of spin up and spin down particles
(see Fig.~\ref{fig:general} for a pictorial illustration).%

\section{Renormalization group analysis}
The resulting effective Hamiltonian is:
\begin{multline} \label{Ham12}
 H=\frac{1}{2\pi} \sum_{a={\rm S,C}} \int \dd x \, v_a K_a \left(\partial_x \varphi_a \right)^2 + \frac{v_a}{K_a} \left(\partial_x \theta_a \right)^2 + \\
+ \int \dd x \, g \left( \mathcal{O}^{\LR}_{2,+1} + \mathcal{O}^{\LR}_{2,-1} \right) + \int \dd x \,h \cos\left(2\sqrt{2}\thetas \right) \, ,
\end{multline}
where we neglected weak terms in the kinetic part which mix spin and charge sectors.
The resulting RG equations in the flow parameter $l$ are~\eqref{reng}:
\begin{align}
 &\partial_l g = g (2- D_g ) \,, \label{reng}\\
 &\partial_l h =h \left( 2-2\Ks \right) +4  \gamma \, g^2 (D_g - \Ks ) \,, \label{renh}
\\ &\partial_l \Ks \propto g^2 D_g (1-\Ks^2) - h^2 \Ks^3 \, , \label{renk}
\end{align}
where $D_g = (\Ks^{-1} + \Ks)/2 + 2 \Kc$ and $\gamma$ is a positive non-universal constant.
Suitable initial conditions are:
$g(l=0) \propto \Omega$, $h(l=0)= \frac{U}{2\pi^2}+ \frac{V}{\pi^2}\cos2\kF$
and $\Ks(l=0)\geq1$, due to repulsive interactions.
In usual spin ladders, where the $\mathcal{O}$ are rapidly oscillating and the $g^2$ correction to Eq.~\eqref{renh} is totally negligible,
$h$ is marginal or irrelevant and is therefore often neglected from the analysis~\cite{giamarchi,nersesyanbook}.
At resonance, however, the presence of the non-oscillating higher harmonics $\mathcal{O}^{\LR}_{2,\pm1}$
generates such a correction, which is positive in the most realistic ranges of the Luttinger parameters
(e.g., for weak repulsive interactions,  $\Kc \lesssim 1$ and $\Ks \gtrsim 1$).
This may change the sign of the initial derivative $\partial_l h$ from negative to positive,
making the term $h$ relevant.
Its RG limit depends on the competition between the scaling dimension of the $\mathcal{O}$ operators in Eq.~\eqref{reng}
and the coefficient of the $g^2$ term in Eq.~\eqref{renh}.%

For onsite interaction only, $1/2 \leq \Kc \leq 1$ and Eq.~\eqref{reng} predicts a suppression of $g$ too rapid to obtain a relevant $h$.
If we instead introduce repulsive interaction $W$ to next-nearest-neighbor sites, $\Kc$ easily reaches such low values that
the coefficient of $g^2$ in Eq.~\eqref{renh} does not get large enough to enhance $h$ either.
Both results are confirmed by our numerical investigations,
where the chiral current signature of the $\nu=1/2$ resonance indeed disappears
for $V \ll U$ and for $W \simeq V$.
If $V \lesssim U$ and $W=0$, instead, the RG function \eqref{renh} becomes positive, making $h$ relevant. In this case $\Ks$ decreases in the renormalization flow \eqref{renk}; if $\Ks$ reaches values smaller than 1, $h$ grows faster than $g$. This determines the opening of a gap in the spin sector of the model (see Sec.~\ref{RGFlow}) originating the resonant state at $\nu=1/2$.

As known from the study of the perturbations of the Moore and Read state~\cite{teokane}, the $h$ coupling can drive the system in a FQH state corresponding to a strongly-paired regime where pairs of fermions merge in bosonic charge 2 objects, which then arrange in a Laughlin $\nu'=1/8$ state. This state is known as the $K=8$ state~\cite{greiter91,wen08}, described often in the context of fully polarized $\nu=1/2$ states.
Similarly to the Laughlin-like resonances, also at $\nu=1/2$ the ladder system mimics the physics of the gapless edge modes in its two-dimensional $K=8$ counterpart. On the qualitative level, its pairing mechanism can be understood by rewriting the spin sector of the model as a function of two massive Majorana modes, whereas the charge sector remains gapless and gives rise to helical modes. The massive Majorana modes imply that the many-body wavefunction acquires a prefactor proportional to ${\rm Pf}({\rm sign}(x_i-x_j)e^{-m_h|x_i-x_j|})$, where $m_h$ is the amplitude of the gap determined by $h$. Such prefactor decays exponentially unless the atoms are arranged in close pairs (see Sec.~\ref{app:correlations}).
In the following we describe the main properties and correlations which define the appearance of this strongly-paired state at the $\nu=1/2$ resonance.

\section{The entanglement properties}
\begin{figure}[ht]
	\centering
	\includegraphics[width=0.33\textwidth]{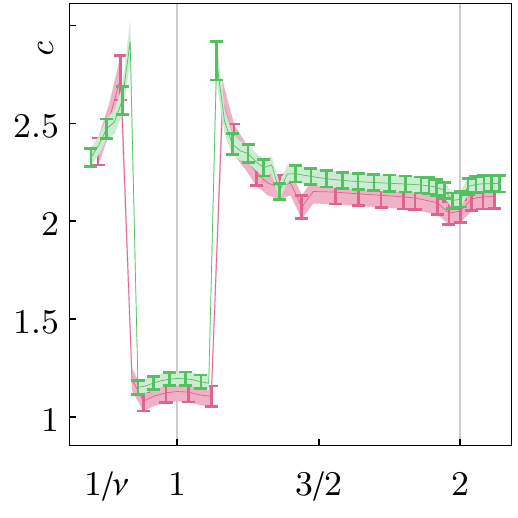}\llap{
  \parbox[b]{2cm}{(a)\\\rule{0ex}{4.5cm}
  }}\hfill
	\includegraphics[width=0.33\textwidth]{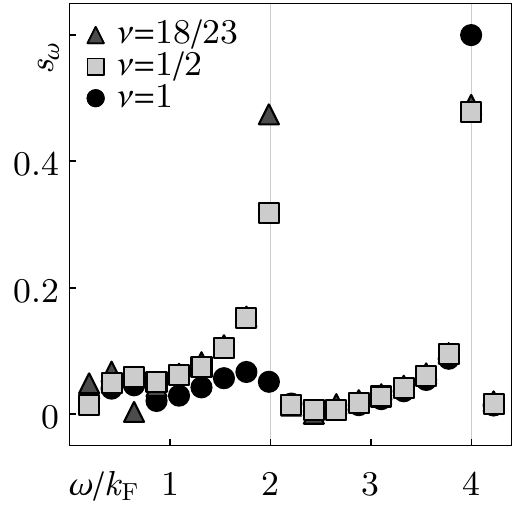}\llap{
  \parbox[b]{2cm}{(b)\\\rule{0ex}{4.5cm}
  }}\hfill
	\includegraphics[width=0.33\textwidth]{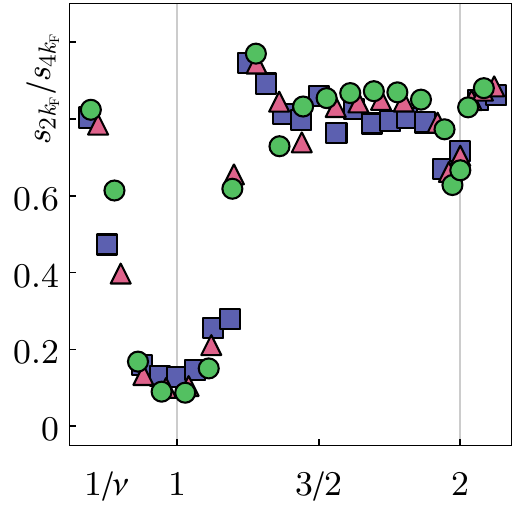}\llap{
  \parbox[b]{2cm}{(c)\\\rule{0ex}{4.5cm}
  }}
 \caption{(a) For $L/N$ = $110/50$ (red), $156/72$ (green) at $\Omega/t=0.02$ and $U/t=V/t=4$, we find a stable $\nu=1/2$ cusp in the central charges surrounded by a flat region.
The other cusp at intermediate fillings is instead a finite-size effect, possibly merging into the $\nu=1$ resonance for $L\to \infty$.
(b) The Fourier spectrum $s_\omega$ of the oscillations $S^\text{OBC}_\ell = S_\ell - S^\text{CFT}_\ell$ reveals two peaks at $2\kF$ and $4\kF$; we show the values $\nu=18/23$ (gray triangle), $1/2$ (gray square) and $1$ (black disk). (c) The $2\kF$ oscillation vanishes for $\nu=1$ and is strongly suppressed at $\nu=1/2$ for the systems of Fig.~\ref{fig:discr}, consistently with the appearance of gaps. For figures of the entanglement entropy, see Fig.~\ref{fig:central_charge_fit}.
 \label{fig:ent}
 }
\end{figure}

To highlight the appearance of energy gaps,
we bipartition the system into segments of lengths $\ell$ and $L-\ell$,
and examine the entanglement spectrum and the associated von Neumann entropy $S_\ell$,
which are readily accessible in MPS simulations~\cite{usualSchollwoeck}.
The entanglement spectrum presents evident discontinuities when driving the system inside and outside the resonances: its gaps change positions around $N_\phi=N/2$ (thus $\nu=1$) and for $N_\phi=N$ ($\nu=1/2$) and different degeneracy patterns appear in these resonances (see Fig.~\ref{fig:entanglement_spectrum}).
Our system is gapless for any value of $\phi$, but the number of helical pairs of gapless modes changes from $c=2$ to $c=1$ inside the resonant states.
This leaves a signature in the entanglement entropy, according, as a first approximation, to the Calabrese and Cardy formula \cite{calabresecardy}, $S_{\ell}^\text{CFT}= c/6 \ln \left[L/\pi \sin \left(\pi \ell/L\right)\right] + S_0$.
For $\nu=1$, the discontinuity of the central charge $c$ from 2 to 1 can be clearly seen close to $\nu=1$ \cite{mazza2016} (Fig. \ref{fig:ent}a).
In the case of the $\nu=1/2$ resonance, instead, we observe only a small downward cusp:
we attribute this to the gap $m_h$ assuming a small value, which in turn determines a correlation length of the gapped sector comparable to the system sizes we were able to investigate (up to $L\simeq150$).
An optimization of $m_h$ goes beyond the scopes of the present work,
but we observe that additional investigations are possible based on algorithms directly tackling the thermodynamic limit~{\cite{mcculloch2012,haegeman2017}}.
We identify a further indication of a gap opening for $\phi=4\kF$
by looking at the corrections induced by open boundary conditions on top of $S_{\ell}^\text{CFT}$
For spinless fermions it is known~\cite{calabrese10} that the entropy $S_\ell$ exhibits
algebraically decaying oscillations with a frequency $2\kF$.
In our model, we observe an additional peak in the frequency spectrum at $4\kF$ (Fig.~\ref{fig:ent}b),
which is suggestive of pairing correlations between the two species \cite{legeza07}.
Noticeably, the oscillations with $2\kF$ disappear at $\nu=1$ and are strongly suppressed at $\nu=1/2$ (Fig.~\ref{fig:ent}c),
thus reenforcing our interpretation of the latter (further detail of our numerical procedures are presented in \ref{sec:error_discussion}  and Fig.~\ref{fig:central_charge_fit}).

\section{The correlations} In order to verify that the two helical modes gapping out at $\nu=1/2$ are indeed the ones in the spin sector,
as in our RG analysis, we examine and compare the following correlation functions:
\begin{align}
C_{\rm s}(r) &\equiv \left\langle \overline{a^\dag_{\Up,r_0} a_{\Dn,r_0} a_{\Up,r_0+r}a^\dag_{\Dn,r_0+r}}\right\rangle_{\rm c} \label{eq:Cs}
\,,\\
C_{\rm p}(r) &\equiv \left\langle \overline{a^\dag_{\Up,r_0} a^\dag_{\Dn,r_0} a_{\Dn,r_0+r}a_{\Up,r_0+r}}\right\rangle_{\rm c} \,, \label{eq:Cp}
\end{align}
where we considered connected two-point correlations averaged over the initial site $r_0$.
\begin{figure}[ht]
	\includegraphics{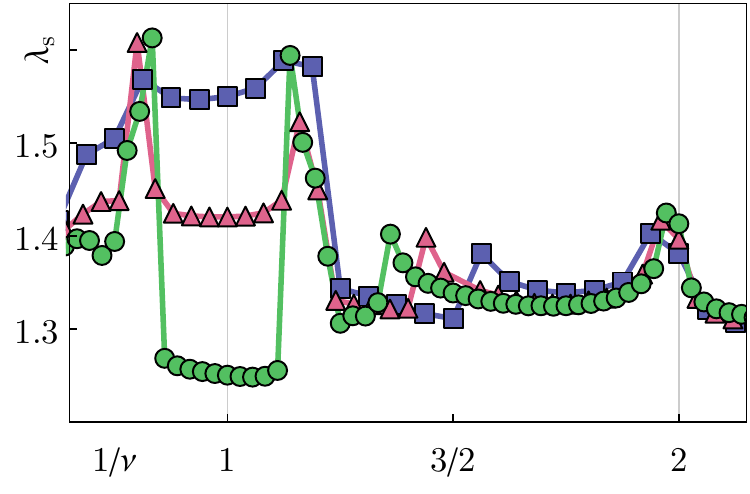}\llap{
  \parbox[b]{3.5cm}{(a)\\\rule{0ex}{3.5cm}
  }}
	\hfill
	\includegraphics{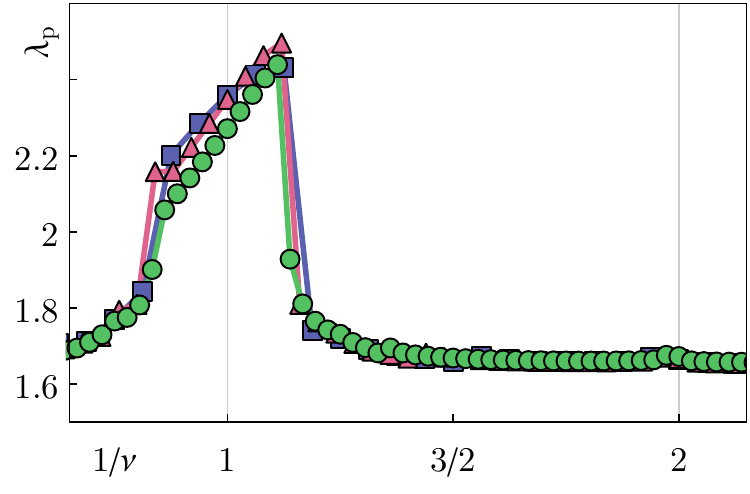}\llap{
  \parbox[b]{3.5cm}{(b)\\\rule{0ex}{3.5cm}
  }}
\caption{
Exponents of the algebraic decay $\lambda_s$ (a) and $\lambda_p$ (b) of $C_s$ and $C_p$
for the setups of Fig.~\ref{fig:discr}.
(a) Around $\nu=1$, $\lambda_s$ decreases as a function of $L$.
At $\nu=1/2$, instead, it displays an increasing cusp, hinting at an exponential decay.
(b) At $\nu=1$, $\lambda_p$ is strongly enhanced, whereas no feature emerges at other fillings:
the tiny cusp at $\nu =1/2$ is an order of magnitude smaller than the one in $\lambda_s$, consistently with a weaker algebraic decay (see Eq.~\ref{eq:corr_s}).
}
\label{fig:corr}
\end{figure}

$C_{\rm s}$ must decay exponentially within the $1/2$ resonance, and algebraically outside it:
its decay is much slower at $\nu=1$, due to the ordering along $\sigma_x$ (Eq.~\eqref{eq:corr_s});
conversely, $C_{\rm p}$ must decay exponentially at the integer resonance,
and algebraically everywhere else, with no distinctive feature at $\nu=1/2$ (Eq.~\eqref{eq:corr_p}).
Due to finite-size limitations, we resort to power-law fits $|C_{\mathrm s/\mathrm p}| \propto r^{-\lambda_{s/p}}$ for these correlations:
In Fig.~\ref{fig:corr}a, we observe that $\lambda_s$ decreases with the system size away from $\nu=1/2$,
i.e., it converges to a true power law in the thermodynamic limit,
while it slightly increases with $L$ when $\nu=1/2$, hinting at a tiny gap opening.
In Fig.~\ref{fig:corr}b, instead, we clearly see $\lambda_p$ strongly enhanced inside the $\nu=1$ resonance,
whereas it remains almost flat and featureless outside it.
The ensemble of signatures is therefore consistent with our RG framework and wavefunction Ansatz.

\section{Conclusions} The study of fermionic FQH state with even denominators has always been more challenging than the odd denominator cases. Here we analyzed the resonant state at $\nu=1/2$ in a spin $1/2$ fermionic chain. With a RG analysis and extensive MPS calculations, we brought compelling evidence that this state is related to the 1D limit of the $K=8$ FQH state~\cite{greiter91,wen08} and it is generated by a gap in the spin sector of the model. The RG technique we adopted constitutes a first step towards the analysis of these effective ladders beyond the semiclassical approximation and can be
extended to the bosonic case at filling $\nu=1$.

Our results are relevant for both ultracold atom ladders in a synthetic dimension~\cite{Celi2014,fallani2015,fallani2016,spielman2015} and nanowires with strong spin-orbit coupling~\cite{depicciotto10,oreg2014,heedt2017}. In the first case the required interactions may be achieved by exploiting dipolar atoms, like Dy, or orbital Feshbach resonances, in the second case electron-electron interactions play a relevant role in the experimental results~\cite{heedt2017} and our tight-binding model can describe their interplay with the Zeeman splitting $\Omega$.

\ack
We gratefully thank L. Mazza for sharing many ideas and for his insightful comments.
We are indebted to A. C. Balram, M. Calvanese Strinati, R. Fazio, K. Le Hur, P. Silvi and L. Taddia for useful discussions,
and to J. J\"unemann for supporting the code for our simulations (partially developed in the framework of the OSCAR grant by DFG to M.R.).
These were performed on the Mogon cluster of the JGU (made available by the CSM and AHRP),
with a code based on a flexible Abelian Symmetric Tensor Networks Library,
developed in collaboration with the group of S. Montangero at Uni-Ulm.
M.B. and M.R. thank the Galileo Galilei Institute of Florence for hospitality during the last stage of this work. M.B. acknowledges Villum Foundation for support. A.H. was supported by a Stufe-I seed grant by JGU to M.R.

\appendix
\section{Conventions for the bosonization of the single-particle Hamiltonian} \label{RG}

We begin by fixing the notation for the bosonization of the single-particle Hamiltonian, starting from the case $\Omega=0$ where the Fermi momenta read $\sigma_z \frac{\phi}{2} \pm k_\mathrm{F}$, with $k_\mathrm{F}=\pi N/(2L)$. We introduce dual bosonic massless field $\varphi$ and $\theta$ such that:
\begin{equation} \label{commutation}
 \left[\theta_{\sigma'}(x'), \varphi_{\sigma}(x) \right]=i \pi \delta_{\sigma\sigma'}\Theta\left(x'-x\right)
\end{equation}
where $\sigma$  refers to the eigenvalues of $\sigma_z$, and $\Theta$ is the Heaviside step function.

Based on these bosonic fields, we can now define fermionic operators which correspond to the harmonics entering in the definition of the fermionic creation and annihilation operators of fermions in the presence of non-linearities of the spectrum. Such operators are:
\begin{align}
 &\psi^{(\mathrm{L},p)}_{\sigma}(x) = \frac{1}{\sqrt{2\pi}}\exp\left[ i\left(   \left(\sigma \frac{\phi}{2} - p k_\mathrm{F} \right)x \right)   - i  p \theta_{\sigma}  (x)+ i \varphi_{\sigma}(x)\right] \,,\\
 &\psi^{(\mathrm{R},p)}_{\sigma}(x) = \frac{1}{\sqrt{2\pi}}\exp\left[ i\left(   \left(\sigma \frac{\phi}{2} + p k_\mathrm{F} \right)x \right)   + i  p \theta_{\sigma}  (x) + i \varphi_{\sigma}(x)\right] \,,
\end{align}
with $p$ a positive odd integer. Based on these vertex operators, the fermionic operators reads:
\begin{equation} \label{a}
 a_{\sigma}(x)= \kappa_\sigma \left[ \sum_{p \; {\rm odd}} \psi^{(\mathrm{L},p)}_{\sigma}(x)
 +\sum_{p \; {\rm odd}} \psi^{(\mathrm{R},p)}_{\sigma}(x)\right]
\end{equation}
where $\kappa_\sigma$ are anticommuting Klein factors.

From these definitions, $\rho_\sigma= \partial_x \theta_\sigma/\pi$ describes the local density of spin $\sigma$ particles close to the Fermi surface and the spin flip operator becomes:
\begin{equation} \label{flip}
 \Omega a^\dag_\Up (x) a_\Dn(x) + {\rm H.c.} = \kappa_\Up \kappa_ \Dn \sum_{\substack{{p,p'}\\{c,c'=\mathrm{L},\mathrm{R}}}} \Psi^{(c,p)\dag}_\Up \Psi^{(c',p')}_\Dn + {\rm H.c.}
\end{equation}

In particular we obtain the following terms:
\begin{multline}
	\mathcal{O}^{\mathrm{L},\mathrm{L}}_{p,p'} \equiv \Psi^{(\mathrm{L},p)\dag}_\Up \Psi^{(\mathrm{L},p')}_\Dn + {\rm H.c.}\\
	\to \kappa_\Up \kappa_ \Dn \exp\left[ i\left(-\phi x + (p-p')k_\mathrm{F} x- \varphi_{\uparrow} + \varphi_{\downarrow} + p\theta_{\uparrow }  - p'\theta_{\downarrow } \right) \right] + {\rm H.c.}  \label{flip0a}
\end{multline}
\begin{multline}
	\mathcal{O}^{\mathrm{R},\mathrm{R}}_{p,p'} \equiv\Psi^{(\mathrm{R},p)\dag}_\Up \Psi^{(\mathrm{R},p')}_\Dn + {\rm H.c.}\\
	\to \kappa_\Up \kappa_ \Dn \exp\left[ i\left( -\phi x - (p-p')k_\mathrm{F} x- \varphi_{\uparrow} + \varphi_{\downarrow} - p\theta_{\uparrow }  + p'\theta_{\downarrow } \right)\right] + {\rm H.c.}  \label{flip0b}
\end{multline}
\begin{multline}
	\mathcal{O}^{\mathrm{L},\mathrm{R}}_{p,p'} \equiv\Psi^{(\mathrm{L},p)\dag}_\Up \Psi^{(\mathrm{R},p')}_\Dn + {\rm H.c.}\\
	\to \kappa_\Up \kappa_ \Dn \exp\left[ i\left( -\phi x + (p+p')k_\mathrm{F} x- \varphi_{\uparrow}  + \varphi_{\downarrow} + p\theta_{\uparrow } + p'\theta_{\downarrow} \right) \right] + {\rm H.c.}\label{flip1}
\end{multline}
\begin{multline}
	\mathcal{O}^{\mathrm{R},\mathrm{L}}_{p,p'} \equiv\Psi^{(\mathrm{R},p)\dag}_\Up \Psi^{(\mathrm{L},p')}_\Dn + {\rm H.c.}\\
	\to \kappa_\Up \kappa_ \Dn \exp\left[ i\left( -\phi x - (p+p')k_\mathrm{F} x- \varphi_{\uparrow}  + \varphi_{\downarrow} - p\theta_{\uparrow } - p'\theta_{\downarrow} \right) \right] + {\rm H.c.}\label{flip2}\,.
\end{multline}

Let us define the charge $\mathrm{C}$ and spin $\mathrm{S}$ fields as:
\begin{equation}
 \varphi_{\mathrm{S}/\mathrm{C}} =\frac{\varphi_\Up \mp \varphi_\Dn}{\sqrt{2}} \,,\quad \theta_{\mathrm{S}/\mathrm{C}} =\frac{\theta_\Up \mp \theta_\Dn}{\sqrt{2}}\,.
\end{equation}
This formulation is useful to evaluate the scaling dimensions of the previous objects. In particular we can introduce the usual Luttinger parameters $K_\mathrm{C}$ and $K_\mathrm{S}$ such that the free part of the Hamiltonian reads:
\begin{equation}
 H=\frac{1}{2\pi} \sum_{a=\mathrm{S},\mathrm{C}} \int dx \, v_a K_a \left(\partial_x \varphi_a \right)^2 + \frac{v_a}{K_a} \left(\partial_x \theta_a \right)^2\,.
\end{equation}
and we define:
\begin{equation}
p_{\mathrm{S}/\mathrm{C}} = \frac{p\mp p'}{2} \,,
\end{equation}
such that $p_\mathrm{C}>0$ and $p_\mathrm{S}=-p_\mathrm{C}+1, \ldots, p_\mathrm{C}-1$. The previous $\mathcal{O}$ operators can be rewritten in terms of the coefficients $p_\mathrm{S}$ and $p_\mathrm{C}$ as shown in the main text for $\mathcal{O}^{\mathrm{L},\mathrm{R}}$. We obtain:
\begin{equation}
\mathcal{O}^{\mathrm{L},\mathrm{R}}_{p_\mathrm{C},p_\mathrm{S}}=\mathcal{O}^{\mathrm{R},\mathrm{L}}_{-p_\mathrm{C},-p_\mathrm{S}}=\mathcal{O}^{\mathrm{L},\mathrm{L}}_{-p_\mathrm{S},p_\mathrm{C}}=\mathcal{O}^{\mathrm{R},\mathrm{R}}_{p_\mathrm{S},-p_\mathrm{C}}\,.
\end{equation}

All the terms with a fast oscillating part in $x$ must be considered as irrelevant in the renormalization group sense. This explains the behavior of the non-interacting system which can be deduced for $p=p'=1$.
In this case the spin flip terms \eqref{flip0a} and \eqref{flip0b} will  disappear due to the fast oscillating term in $\phi x$ if $\phi \neq 0$. The terms \eqref{flip1} and \eqref{flip2} loose their fast oscillating behavior at the resonances $\phi=\pm 2k_\mathrm{F}$ respectively. This explains the fact that the Zeeman term gaps two of the four gapless modes when the chemical potential is tuned at $\mu=2\cos(\phi/2)$. In this way the system remains with two gapless helical modes and, interpreting the spin as a synthetic dimension, we can consider the resonance $\phi =\pm 2k_\mathrm{F}$ as a one-dimensional limit of the integer quantum Hall state at filling $\nu=1$. In particular we define the filling as:
\begin{equation}
 \nu=\frac{1}{p_\mathrm{C}}=\frac{2k_\mathrm{F}}{\phi }=\frac{\pi N}{\phi L}\,,
\end{equation}
which corresponds to the ratio between number of particles and magnetic fluxes as in the usual quantum Hall setups.
We also observe that there exist a special point at $\phi =\pi$ where, for the non-interacting case $p=p'=1$ both \eqref{flip1} and \eqref{flip2} become relevant at $N=L$. This point has been extensively studied in \cite{barbarino15}.

Let us mention that, for $\phi =2k_\mathrm{F}$, not only $\mathcal{O}^{\mathrm{L},\mathrm{R}}_{p_\mathrm{C}=1,p_\mathrm{S}=0}$ looses its fast oscillating dependence, but, in general, all the operators $\mathcal{O}^{\mathrm{L},\mathrm{L}}_{p_\mathrm{C}\geq 2,p_\mathrm{S}=1}$ and $\mathcal{O}^{\mathrm{R},\mathrm{R}}_{p_\mathrm{C}\geq 2,p_\mathrm{S}=-1}$. Due to the coefficients of the bosonic fields, though, these additional operators are less relevant because of the higher scaling dimension. This is a common feature: at a given resonance with filling $\nu$, the resonant operators $\mathcal{O}^{\mathrm{L},\mathrm{R}}_{p_\mathrm{C},p_\mathrm{S}}$ or $\mathcal{O}^{\mathrm{R},\mathrm{L}}_{p_\mathrm{C},p_\mathrm{S}}$ will be more relevant than the resonant operators $\mathcal{O}^{\mathrm{L},\mathrm{L}}_{p_\mathrm{C}',p_\mathrm{S}'=p_\mathrm{C}}$ and $\mathcal{O}^{\mathrm{R},\mathrm{R}}_{p_\mathrm{C}',p_\mathrm{S}'=p_\mathrm{C}}$, because, in general $p_\mathrm{C}' > p_\mathrm{S}'=p_\mathrm{C}>|p_\mathrm{S}|$ and the scaling dimension of the operators $\mathcal{O}$ reads:
\begin{align}
\mathcal{O}^{\mathrm{L},\mathrm{R}}_{p_\mathrm{C},p_\mathrm{S}}, \,\mathcal{O}^{\mathrm{R},\mathrm{L}}_{p_\mathrm{C},p_\mathrm{S}}  \;&\rightarrow \; D_g = \frac{1}{2}\left[\frac{1}{K_\mathrm{S}} + p_\mathrm{S}^2 K_\mathrm{S} +p_\mathrm{C}^2 K_\mathrm{C}\right]\,,\\
\mathcal{O}^{\mathrm{L},\mathrm{L}}_{p'_\mathrm{C},p'_\mathrm{S}},\, \mathcal{O}^{\mathrm{R},\mathrm{R}}_{p'_\mathrm{C},p'_\mathrm{S}}  \;&\rightarrow \; D'_g = \frac{1}{2}\left[\frac{1}{K_\mathrm{S}} + p^{\prime 2}_\mathrm{C} K_\mathrm{S} +p^{\prime 2}_\mathrm{S} K_\mathrm{C}\right]\,.
\end{align}

\section{The renormalization group equations at the $\nu=1/2$ resonance}\label{RGFlow}
Our scope is to characterize the ground state of the system in proximity of the $\phi =\pm 4k_\mathrm{F}$ resonances. First we observe that the nature of these resonances do not depend on the sign of $\phi $. The Hamiltonian is indeed invariant by the simultaneous transformation $\phi \to -\phi $ and $\Up \leftrightarrow \Dn$. Therefore we simply discuss the case with $\phi =2k_\mathrm{F}$.

As observed in the main text, at this resonance there are two competing operators which are generated by the Zeeman term and loose their fast oscillating dependence:
\begin{align}
 &\mathcal{O}^{\mathrm{L},\mathrm{R}}_{p_\mathrm{C}=2,p_\mathrm{S}=+1} = \kappa_\Up \kappa_ \Dn \exp\left[ i\left(- \varphi_{\uparrow}  + \varphi_{\downarrow} + \theta_{\uparrow } + 3\theta_{\downarrow} \right)\right] + {\rm H.c.} \label{O13} \\
 &\mathcal{O}^{\mathrm{L},\mathrm{R}}_{p_\mathrm{C}=2,p_\mathrm{S}=-1} = \kappa_\Up \kappa_ \Dn \exp\left[ i\left(- \varphi_{\uparrow}  + \varphi_{\downarrow} + 3\theta_{\uparrow } + \theta_{\downarrow} \right)\right] + {\rm H.c.}\,. \label{O31}
\end{align}
When the time-reversal symmetry is preserved by the interaction, these operators share the same amplitude and scaling dimension. As in the general case, also other terms will loose their fast oscillating behavior at this resonance, the most relevant being $\mathcal{O}^{\mathrm{L},\mathrm{L}}_{3,+2}$ and $\mathcal{O}^{\mathrm{R},\mathrm{R}}_{3,-2}$. $\mathcal{O}^{\mathrm{L},\mathrm{R}}_{2,-1}$ and $\mathcal{O}^{\mathrm{L},\mathrm{R}}_{2,+1}$, however, are the most relevant and we will neglect the others in our analysis.

The main property of this resonance, characterized by an even denominator, is that the operators \eqref{O13} and \eqref{O31} do not commute with each other and they share the same scaling dimension and amplitude. Therefore it is not a priori clear what is the mechanism that is able to open a gap in a pair or gapless modes and a more refined renormalization group analysis is needed.

Let us consider the following interaction as a perturbation of the free bosonic Hamiltonian:
\begin{equation} \label{HI}
 H_I = g  \int dx \,  \left( \mathcal{O}^{\mathrm{L},\mathrm{R}}_{2,+1}  + \mathcal{O}^{\mathrm{L},\mathrm{R}}_{2,-1}\right)
 = -2g i\kappa_\Up \kappa_ \Dn \int dx  \sin\left(\varphi_{\downarrow}-\varphi_{\uparrow } + 2\theta_{\uparrow} + 2\theta_{\downarrow} \right) \sin\left( \theta_{\uparrow} - \theta_{\downarrow}  \right)
\end{equation}
Here the combination of fields inside the sines in the last term do not commute as well and, also in this formulation, it is hard to understand what is the mechanism determining a gap. The sine contributions take into account on one side the Campbell-Baker-Hausdorff formula and, on the other, the signs determined by the Klein factors.

A similar situation has already been analyzed in works related to the interactions in fermionic ladder systems. In particular $H_I$ appears as a higher harmonic term of the tunneling interactions considered in \cite{yakovenko92} and \cite{nersesyan93}, whose renormalization analysis bring to the appearance of pair-tunneling terms in the Hamiltonian.

In the following we will examine the interaction $H_I$ under the light of the Wilsonian renormalization method and we will determine the most relevant operators generated in the scaling flow of \eqref{HI}. We will follow the renormalization techniques described in \cite{nersesyanbook}.

To this purpose let us introduce the following notation for the interaction:
\begin{equation} \label{HII}
 H_I = g \kappa_\Up \kappa_ \Dn \int dx \sum_{\mu, \nu = \pm 1} \mu e^{i\mu \Phi_{2,-1}} + \nu e^{i\nu \Phi_{2,+1}}\,,
\end{equation}
where the fields $\Phi_{2,\pm 1}$ are defined in the main text and the signs of the vertex operators in \eqref{HII} account for the Klein factors.

The action (in Euclidean space) of the system can be written as:
\begin{equation}
 S = \frac{1}{2\pi} \int d^2 x \left[ \sum_a \frac{K_a}{v_a} (\partial_t \varphi_a)^2 + K_a v_a (\partial_x \varphi_a)^2 \right] + g \kappa_\Up \kappa_ \Dn \int d^2x \sum_{\mu, \nu = \pm 1} \mu e^{i\mu \Phi_{2,-1}} + \nu e^{i\nu \Phi_{2,+1}}\,,
\end{equation}
where the first term constitutes the free bosonic action $S_0$ and the second will be a perturbation $S_I$.

The idea behind the Wilsonian renormalization group is to consider a cutoff in momentum $\Lambda$ and a small scale parameter $l$ such that we can rescale the cutoff to a smaller value $\Lambda'=\Lambda e^{-l}$ corresponding to $\Lambda'/\Lambda \approx 1-dl$. Following the standard approach we can split the field $\varphi$ and its dual $\theta$ into a ``slow'' and a ``fast'' component. The first includes all the momenta smaller than $\Lambda'$, the second the momenta included in the shell $\Lambda'<k<\Lambda$:
\begin{equation}
 \varphi_a(x,t)= \varphi_{{\sf s},a}(x,t)+\varphi_{{\sf f},a}(x,t)\,,\quad  \theta_a(x,t)= \theta_{{\sf s},a}(x,t)+\theta_{{\sf f},a}(x,t)\,.
\end{equation}
Analogously also the operators $\Phi$ will be decomposed in $\Phi_{\sf s} + \Phi_{\sf f}$.
To understand the renormalization flow, we must derive an effective action for the slow modes only, by averaging over the fast modes. One obtains:
\begin{multline} \label{seff}
 S_{\rm eff}(\Lambda')=S_0(\varphi_{\sf s}) - \ln \left\langle e^{-S_I(\varphi_{\sf s}+\varphi_{\sf f})} \right\rangle_{\sf f}\\
 \approx S_0(\varphi_{\sf s}) + \underbrace{\left\langle S_I(\varphi_{\sf s}+\varphi_{\sf f})\right\rangle_{\sf f}}_{\mathcal{A}} -\frac{1}{2} \left(\underbrace{\left\langle S_I^2(\varphi_{\sf s}+\varphi_{\sf f}) \right\rangle_{\sf f}}_{\mathcal{B}} - \underbrace{\left\langle S_I(\varphi_{\sf s}+\varphi_{\sf f}) \right\rangle^2_{\sf f}}_{\mathcal{A}^2} \right) + \ldots
\end{multline}
In this expression, the average values are taken integrating over the fast modes. In the following we will evaluate the values of $\mathcal{A}$ and $\mathcal{B}$ to obtain $S_{\rm eff}$. It is useful to consider the following relations:
\begin{multline}
 \left\langle e^{ i \sum_j a_j \varphi_{\sf f}(x_j) + i \sum_k b_k \theta_{\sf f}(x_k)}\right\rangle_{\sf f} = \exp\left[ -\frac{\sum_j a_j^2}{2}\left\langle \varphi_{\sf f}^2 \right\rangle_{\sf f} - \frac{\sum_k b_k^2}{2}\left\langle \theta_{\sf f}^2 \right\rangle_{\sf f}\right.
 \\\left. - \sum_{j < j'} a_ja_{j'}\left\langle \varphi_{\sf f}(x_j) \varphi_{\sf f}(x_{j'})\right\rangle_{\sf f} - \sum_{k < k'} b_kb_{k'}\left\langle \theta_{\sf f}(x_k) \theta_{\sf f}(x_{k'})\right\rangle_{\sf f}\right]  \,, \label{correlation1}
\end{multline}
where the scaling of the correlation functions is given by:
\begin{align}
 &\left\langle \varphi_{\sf f}^2(x) \right\rangle_{\sf f} = \int_{\Lambda' < k < \Lambda} \frac{d^2k}{4\pi} \frac{1}{Kk^2}=\int_{\Lambda'}^\Lambda \frac{dk}{2} \frac{1}{Kk}= \frac{1}{2K}\ln\frac{\Lambda}{\Lambda'}\,, \\
 &\left\langle \varphi_{\sf f}(x,t) \varphi_{\sf f}(x',t') \right\rangle_{\sf f} =  \int_{\Lambda'}^{\Lambda} \frac{dk}{2} \frac{1}{Kk} e^{ikr} \approx \frac{C(r)}{2K} \ln\frac{\Lambda}{\Lambda'}
\end{align}
where the logarithm captures the scaling behavior, and $C(r)$ is a short-range function of $r=\sqrt{(t-t')^2+(x-x')^2}$ (to be more precise, in this case a sharp cutoff, $C$ is not really short-ranged, but it can be made short-ranged with better cutoffs \cite{kogut}). Analogously we have:
\begin{equation}
 \left\langle \theta_{\sf f}^2(x) \right\rangle_{\sf f} = \frac{K}{2}\ln\frac{\Lambda}{\Lambda'}\,,\quad \left\langle \theta_{\sf f}(x,t) \theta_{\sf f}(x',t') \right\rangle_{\sf f} \approx \frac{C(r)K}{2} \ln\frac{\Lambda}{\Lambda'}\,.
\end{equation}
Let us now calculate the first-order term $\mathcal{A}$:
\begin{multline}
 \mathcal{A}=g \kappa_\Up \kappa_ \Dn \sum_{\mu, \nu = \pm 1}\int d^2x  \left\langle \mu e^{i\mu \Phi_{2,-1}}\right\rangle_{\sf f}  + \left\langle \nu e^{i\nu \Phi_{2,+1}}\right\rangle_{\sf f}= \\
 =g \kappa_\Up \kappa_ \Dn \sum_{\mu, \nu = \pm 1} \int d^2x  \mu e^{i\mu \Phi_{2,-1,{\sf s}}} \left\langle e^{i\mu \Phi_{2,-1,{\sf f}}}\right\rangle_{\sf f} + \nu e^{i\nu \Phi_{2,+1,{\sf s}}} \left\langle e^{i\nu \Phi_{2,+1,{\sf f}}}\right\rangle_{\sf f} = \\
 = g \kappa_\Up \kappa_ \Dn \left( \frac{\Lambda}{\Lambda'}\right)^{-\left(2K_\mathrm{C}+\frac{K_\mathrm{S}}{2} +\frac{1}{2K_\mathrm{S}} \right) } \sum_{\mu, \nu = \pm 1}\int d^2x \mu e^{i\mu \Phi_{2,-1,{\sf s}}} + \nu e^{i\nu \Phi_{2,+1,{\sf s}}}
\end{multline}
where we used the previous expressions for the correlations of the vertex operators and one has to separately consider the two spin species. The scaling dimension $D_g$ appears in the scaling of the cutoff factor.
We complete the renormalization with a final rescaling of the space coordinates, $d^2x = d^2x'(1+2dl)$; we obtain:
\begin{equation}
 \mathcal{A} \approx \left(1+2dl -D_g dl \right) S_I(\varphi_{\sf s})\,,
\end{equation}
which provides the renormalization at first order of the interaction. As expected from the scaling argument we obtain:
\begin{equation}
 \frac{dg}{dl}=g \left[ 2-D_g\right] \,.
\end{equation}
It is now easy to write also the operator $\mathcal{A}^2$:
\begin{multline} \label{A2}
 \mathcal{A}^2 = -g^2\left(1+4dl -2D_gdl \right)
 \sum_{\mu_1,\mu_2,\nu_1,\nu_2 = \pm 1} \int d^2x'_1 d^2x'_2 \left(\mu_1 e^{i\mu_1 \Phi_{2,-1,{\sf s}}(x_1')} + \nu_1 e^{i\nu_1 \Phi_{2,+1,{\sf s}}(x_1')} \right)\cdot \\
 \left(\mu_2 e^{i\mu_2 \Phi_{2,-1,{\sf s}}(x_2')} + \nu_2 e^{i\nu_2 \Phi_{2,+1,{\sf s}}(x_2')} \right) \,,
\end{multline}
here the Klein factors have been squared away, leaving a minus sign due to anticommutation. To complete the second order description of $S_{\rm eff}$ we now calculate the term $\mathcal{B}$ in \eqref{seff}:
\begin{align}
 \mathcal{B}=-g^2 \sum_{\mu_1,\mu_2,\nu_1,\nu_2 = \pm 1} &\int d^2x_1 d^2x_2 \left\langle \left(\mu_1 e^{i\mu_1 \Phi_{2,-1}(x_1)} + \nu_1 e^{i\nu_1 \Phi_{2,+1}(x_1)} \right)\right.\nonumber\\&\phantom{\int d^2x_1 d^2x_2\left\langle \right.}\left.\left(\mu_2 e^{i\mu_2 \Phi_{2,-1}(x_2)} + \nu_2 e^{i\nu_2 \Phi_{2,+1}(x_2)} \right)\right\rangle_{\sf f}
 \end{align}
 \begin{align}
 \mathcal B= -g^2 \sum_{\mu_1,\mu_2,\nu_1,\nu_2 = \pm 1} &\int d^2x_1 d^2x_2 \Big [  \nonumber \\
 &-\mu_1 \mu_2 e^{i\mu_1 \Phi_{2,-1,{\sf s}}(x_1)+i\mu_2 \Phi_{2,-1,{\sf s}}(x_2)} \left\langle e^{i\mu_1 \Phi_{2,-1,{\sf f}}(x_1)+i\mu_2 \Phi_{2,-1,{\sf f}}(x_2)}\right\rangle_{\sf f} + \label{mumu}\\
 &-\nu_1 \nu_2 e^{i\nu_1 \Phi_{2,+1,{\sf s}}(x_1)+i\nu_2 \Phi_{2,+1,{\sf s}}(x_2)} \left\langle e^{i\nu_1 \Phi_{2,+1,{\sf f}}(x_1)+i\nu_2 \Phi_{2,+1,{\sf f}}(x_2)}\right\rangle_{\sf f} +  \label{nunu}\\
 &-\mu_1 \nu_2 e^{i\mu_1 \Phi_{2,-1,{\sf s}}(x_1)+i\nu_2 \Phi_{2,+1,{\sf s}}(x_2)} \left\langle e^{i\mu_1 \Phi_{2,-1,{\sf f}}(x_1)+i\nu_2 \Phi_{2,+1,{\sf f}}(x_2)}\right\rangle_{\sf f} + \label{munu}\\
 &-\nu_1 \mu_2 e^{i\nu_1 \Phi_{2,+1,{\sf s}}(x_1)+i\mu_2 \Phi_{2,-1,{\sf s}}(x_2)} \left\langle e^{i\nu_1 \Phi_{2,+1,{\sf f}}(x_1)+i\mu_2 \Phi_{2,-1,{\sf f}}(x_2)}\right\rangle_{\sf f} \Big ]\,.\label{numu}
\end{align}
Here the signs are determined by the commutation relations of the $\kappa$ operators and by the Campbell-Baker-Hausdorff formula. We now apply \eqref{correlation1} and the definition of the correlation functions of the bosonic fields to evaluate the scaling of the average values appearing in (\ref{mumu}-\ref{numu}). We obtain:
\begin{align}
 \left\langle e^{i\mu_1 \Phi_{2,-1,{\sf f}}(x_1)+i\mu_2 \Phi_{2,-1,{\sf f}}(x_2)}\right\rangle_{\sf f} =& \, \left( \frac{\Lambda}{\Lambda'}\right)^{-2D_g}\left( \frac{\Lambda}{\Lambda'}\right)^{-C(x_1-x_2)\mu_1 \mu_2 2D_g}\,, \\
 \left\langle e^{i\nu_1 \Phi_{2,+1,{\sf f}}(x_1)+i\nu_2 \Phi_{2,+1,{\sf f}}(x_2)}\right\rangle_{\sf f}=& \, \left( \frac{\Lambda}{\Lambda'}\right)^{-2D_g}\left( \frac{\Lambda}{\Lambda'}\right)^{-C(x_1-x_2)\nu_1 \nu_2 2D_g}\,, \\
 \left\langle e^{i\mu_1 \Phi_{2,-1,{\sf f}}(x_1)+i\nu_2 \Phi_{2,+1,{\sf f}}(x_2)}\right\rangle_{\sf f} =& \, \left( \frac{\Lambda}{\Lambda'}\right)^{-2D_g}\left( \frac{\Lambda}{\Lambda'}\right)^{-C(x_1-x_2)\mu_1 \nu_2\left(4K_\mathrm{C}-K_\mathrm{S}+\frac{1}{K_\mathrm{S}}\right)}\,, \\
 \left\langle e^{i\nu_1 \Phi_{2,+1,{\sf f}}(x_1)+i\mu_2 \Phi_{2,-1,{\sf f}}(x_2)}\right\rangle_{\sf f} =& \, \left( \frac{\Lambda}{\Lambda'}\right)^{-2D_g}\left( \frac{\Lambda}{\Lambda'}\right)^{-C(x_1-x_2)\nu_1 \mu_2\left(4K_\mathrm{C}-K_\mathrm{S}+\frac{1}{K_\mathrm{S}}\right)}\,.
\end{align}
In these equations $C(x_1-x_2)$ is a non-universal function appearing from the two-point correlation functions which depends on the kind of cutoff but can be considered short-ranged (see \cite{nersesyanbook,kogut} for a more detailed discussion). Rewriting these scaling terms as a function of $dl$ and considering also the scaling of the real space coordinates, we can approximate $\mathcal{B}$ as:
\begin{align}
 \mathcal{B} \approx -g^2&\left(1+4dl- 2D_g dl\right)  \sum_{\mu_1,\mu_2,\nu_1,\nu_2 = \pm 1} \int d^2x'_1 d^2x'_2 \Big [  \nonumber \\
 &\left(1 -\mu_1 \mu_2 2D_g C(x_1'-x_2')dl\right)\mu_1 \mu_2 e^{i\mu_1 \Phi_{2,-1,{\sf s}}(x'_1)}e^{i\mu_2 \Phi_{2,-1,{\sf s}}(x'_2)}  + \label{mumu2}\\
 &\left(1 -\nu_1 \nu_2 2D_g C(x_1'-x_2')dl\right)\nu_1 \nu_2 e^{i\nu_1 \Phi_{2,+1,{\sf s}}(x'_1)}e^{i\nu_2 \Phi_{2,+1,{\sf s}}(x'_2)}  +  \label{nunu2}\\
 &\left(1 -\mu_1 \nu_2\left(4K_\mathrm{C}-K_\mathrm{S}+\frac{1}{K_\mathrm{S}}\right)C(x_1'-x_2')dl\right)\mu_1 \nu_2 e^{i\mu_1 \Phi_{2,-1,{\sf s}}(x'_1)}e^{i\nu_2 \Phi_{2,+1,{\sf s}}(x'_2)}  + \label{munu2}\\
 &\left(1 -\nu_1 \mu_2\left(4K_\mathrm{C}-K_\mathrm{S}+\frac{1}{K_\mathrm{S}}\right)C(x_1'-x_2')dl\right)\nu_1 \mu_2 e^{i\nu_1 \Phi_{2,+1,{\sf s}}(x'_1)}e^{i\mu_2 \Phi_{2,-1,{\sf s}}(x'_2)}  \Big ]\,.\label{numu2}
\end{align}
We observe that, in this expression, the terms independent on the two-point correlation functions (thus not proportional to the function $C$) coincide, as expected, with $\mathcal{A}^2$ in \eqref{A2} and they erase in $S_{\rm eff}$.

We can now write the second-order interaction terms of $S_{\rm eff}$:
\begin{align}
 S_{I, {\rm eff}}(\varphi_{\sf s}) = &\mathcal{A} -\frac{1}{2}\left(\mathcal{B} - \mathcal{A}^2 \right) = \nonumber \\
 =& \left(1+2dl -dl D_g\right) S_I(\varphi_{\sf s}) - \frac{g^2dl}{2} \sum_{\mu_1,\mu_2,\nu_1,\nu_2 = \pm 1} \int d^2x'_1 d^2x'_2 C(x_1'-x_2')\Big[  \nonumber \\
 &2D_g e^{i\mu_1 \Phi_{2,-1,{\sf s}}(x'_1)}e^{i\mu_2 \Phi_{2,-1,{\sf s}}(x'_2)} + 2D_g e^{i\nu_1 \Phi_{2,+1,{\sf s}}(x'_1)}e^{i\nu_2 \Phi_{2,+1,{\sf s}}(x'_2)} + \label{mumu3}\\
 & + \left(4K_\mathrm{C}-K_\mathrm{S}+\frac{1}{K_\mathrm{S}}\right) e^{i\mu_1 \Phi_{2,-1,{\sf s}}(x'_1)}e^{i\nu_2 \Phi_{2,+1,{\sf s}}(x'_2)} \nonumber\\
 & +\left(4K_\mathrm{C}-K_\mathrm{S}+\frac{1}{K_\mathrm{S}}\right) e^{i\nu_1 \Phi_{2,+1,{\sf s}}(x'_1)}e^{i\mu_2 \Phi_{2,-1,{\sf s}}(x'_2)} \Big ]\,. \label{munu3}
\end{align}
Let us consider first the terms in line \eqref{mumu3}. It is helpful to distinguish the following cases: (i) $\mu_1=\mu_2$ and $\nu_1=\nu_2$; (ii)
$\mu_1=-\mu_2$ and $\nu_1=-\nu_2$.

In case (i) the resulting operators are vertex operators proportional to $\exp\left[\pm i\left(  \Phi_{p_\mathrm{C},p_\mathrm{S}}(x_1') + \Phi_{p_\mathrm{C},p_\mathrm{S}}(x_2') \right) \right]$. In this expression it is convenient to exploit the fact that the function $C$ is short ranged, such that we can distinguish a center of mass coordinate and a relative coordinate. In particular, to evaluate the terms of the kind (i), we may approximate $C=c\delta\left( x'_1-x'_2\right) $ where $c$ is a non-universal constant (following \cite{nersesyanbook,kogut,giamarchi}, in the case of sharp cutoff, it can be expressed in term of integrals of the Bessel function). A more refined discussion can be found in \cite{wiegmann78,kogut}. These terms thus generate interactions of the kind $\exp\left[\pm i\left(  2\Phi_{p_\mathrm{C},p_\mathrm{S}}(x') \right) \right]$ which are characterized by a large scaling dimension, $4D_g$, thus they are irrelevant and we neglect them.

In case (ii) instead, we obtain vertex operators of the difference of the fields $\Phi$; they are of the form:
\begin{equation} \label{kin}
 -\sum_{s=\pm1}\exp\left[i s \left(  \Phi_{p_\mathrm{C},p_\mathrm{S}}(x_1') - \Phi_{p_\mathrm{C},p_\mathrm{S}}(x_2') \right) \right] \approx -2\cos\left(  \alpha  \partial_x\Phi_{p_\mathrm{C},p_\mathrm{S}}(x) \right) \approx -2 + \alpha^2 (\partial_x \Phi_{p_\mathrm{C},p_\mathrm{S}})^2
\end{equation}
where we exploited $C$ being short range and we introduced an effective non-universal range $\alpha$ determined by $C$ \cite{nersesyanbook,kogut}. We observe that, similarly to the sine-Gordon model, the terms \eqref{kin} contribute to the renormalization of the free action $S_0$, thus of $K$ and $v$ such that:
\begin{equation} \label{kin2}
 S_0 \to S_0 + \int d^2x' {g^2} \alpha^2 D_g dl \left[(\partial_x \Phi_{2,-1})^2  + (\partial_x \Phi_{2,+1})^2\right]\,.
\end{equation}
This additional kinetic term is not diagonal in $\varphi$ and $\theta$, and, in principle, this spoils the possibility of separating the spin and charge degrees of freedom. This separation, however, is violated only by the term in $\partial_x \varphi_\mathrm{S} \partial_x\theta_\mathrm{C}$. To derive the flow of the Luttinger parameters, we will neglect this violation and we will consider only the usual terms in $(\partial_x \theta)^2$ and $(\partial_x \varphi)^2$.

Let us consider now, instead, the terms in line \eqref{munu3}. Also in this case it is convenient to distinguish the cases (i), with $\mu_1=\nu_2$ and $\mu_2=\nu_1$, and (ii), with $\mu_1=-\nu_2$ and $\mu_2=-\nu_1$. Concerning case (i), when we consider the short range constraint imposed by $C$, these terms generate operators like $\exp\left[\pm i\left(\Phi_{2,-1} + \Phi_{2,+1} \right) \right]$ whose scaling dimension is $2\left(\frac{1}{K_\mathrm{S}} +4K_\mathrm{C} \right)$. As before, we neglect these terms because they are always irrelevant.

In case (ii), instead, by approximating $C=\gamma\delta\left( x'_1-x'_2\right) $, we obtain terms of the kind:
\begin{multline} \label{pairingint}
 -2\gamma\left(4K_\mathrm{C}-K_\mathrm{S}+\frac{1}{K_\mathrm{S}}\right)\sum_{s=\pm1} \exp\left[\pm i\left(\Phi_{2,-1} - \Phi_{2,+1} \right) \right]\\
 = -4\gamma\left(4K_\mathrm{C}-K_\mathrm{S}+\frac{1}{K_\mathrm{S}}\right) \cos\left(2\theta_\Up -2 \theta_\Dn \right) \,;
\end{multline}
here the additional minus sign comes from the Campbell-Baker-Haussdorf formula.
Therefore the $h$ interaction term, with scaling dimension $2K_\mathrm{S}$ emerges, and it is relevant for any $K_\mathrm{S}<1$. This term coincides with a four body operator given by the product of backscattering terms of the two spin species, as sketched in Fig. 1 of the main text. The final expression for $S_{\rm eff}$ becomes:
\begin{multline}
 S_{\rm eff} = S_0 + \int d^2x' {g^2} \alpha^2 D_g dl \left[(\partial_x \Phi_{2,-1})^2  + (\partial_x \Phi_{2,+1})^2\right]  + \\
 + g \kappa_\Up \kappa_\Dn \left(1+2dl -D_g dl \right)  \int d^2x' \left[ \sum_{\mu, \nu = \pm 1} \mu e^{i\mu \Phi_{2,-1}} + \nu e^{i\nu \Phi_{2,+1}} \right] + \\
 +2\gamma g^2\left(4K_\mathrm{C}-K_\mathrm{S}+\frac{1}{K_\mathrm{S}}\right)dl \int d^2x'\cos\left(2\theta_\Up -2 \theta_\Dn \right)\,.
\end{multline}
Therefore, following the approach in \cite{yakovenko92,nersesyanbook}, to study the relevant terms of the renormalization at second order, we must consider the following interaction part for the Lagrangian:
\begin{equation}
 \mathcal{L}_I = \int dx \, g \left( \mathcal{O}^{\mathrm{L},\mathrm{R}}_{2,-1}(x) + \mathcal{O}^{\mathrm{L},\mathrm{R}}_{2,+1}(x) \right) + \int dx \,h \cos\left(2\theta_\Up -2 \theta_\Dn \right)
\end{equation}
where $g$ and $h$ are coupling constants flowing with the renormalization scale $l$ such that they fulfill the RG equations:
\begin{align}
 &\frac{dg}{dl} = g \left[ 2-2K_\mathrm{C}-\frac{K_\mathrm{S}}{2}-\frac{1}{2K_\mathrm{S}}\right] \,, \label{reng}\\
 &\frac{dh}{dl} =h \left( 2-2K_\mathrm{S} \right) +4\gamma g^2\left(2K_\mathrm{C}-\frac{K_\mathrm{S}}{2}+\frac{1}{2K_\mathrm{S}} \right) \,, \label{renh}
\end{align}
where the boundary conditions are given by $g(l=0)=\Omega/(2\pi)$ and $h(l=0)= \frac{U}{2\pi^2}+ \frac{V}{\pi^2}\cos2k_\mathrm{F}$ (see Sec. \ref{sec:initial}).

Besides these equations, it is possible to derive RG equations for the parameters $K$ and $v$ from Eq. \eqref{kin2}. Since we are mostly interested in the scaling of $K_\mathrm{S}$ we consider its behavior as a function of $g^2$. From Eq. \eqref{kin2} we get:
\begin{align}
K'_\mathrm{S}v'_\mathrm{S} &= K_\mathrm{S} v_\mathrm{S} + 8 \pi g^2 \alpha^2 \left(2K_\mathrm{C} + \frac{K_\mathrm{S}}{2}+\frac{1}{2K_\mathrm{S}}\right) dl \,, \\
\frac{v_\mathrm{S}'}{K_\mathrm{S}'} &= \frac{v_\mathrm{S}}{K_\mathrm{S}} + 8 \pi g^2 \alpha^2 \left(2K_\mathrm{C} + \frac{K_\mathrm{S}}{2}+\frac{1}{2K_\mathrm{S}}\right) dl\,;
\end{align}
these equations imply:
\begin{equation} \label{renk}
\frac{dK_\mathrm{S}}{dl}= \frac{4\pi g^2 \alpha^2}{v_\mathrm{S}}\left(2K_\mathrm{C} + \frac{K_\mathrm{S}}{2}+\frac{1}{2K_\mathrm{S}}\right)\left(1-K_\mathrm{S}^2\right) - \frac{4\pi h^2 \alpha^2}{v_\mathrm{S}}K_\mathrm{S}^3
\end{equation}
where the second term is derived from the RG analysis of the $h$ term, following, for example, \cite{giamarchi}. We observe that, if there exists a fixed point with finite values for $g$ and $h$, then the fixed value of $K_\mathrm{S}$ in this fixed point would be $\tilde{K}_\mathrm{S}<1$.
To qualitatively understand the behavior of the renormalization flow, we thus substitute the flowing $K_\mathrm{S}$ with its fixed point value $\tilde{K}_\mathrm{S}<1$. Hence, neglecting the flow of $K_\mathrm{S}$, the solution of \eqref{reng} and \eqref{renh} is:
\begin{align}
 g(l) \propto & \,\Omega e^{\left(2-2K_\mathrm{C}-\frac{\tilde{K}_\mathrm{S}}{2}-\frac{1}{2\tilde{K}_\mathrm{S}} \right)l }\,,\\
 h(l) = &  (h(0)+\beta)e^{(2-2\tilde{K}_\mathrm{S})l} -\beta e^{\left(4-\frac{1}{\tilde{K}_\mathrm{S}}-4K_\mathrm{C} -\tilde{K}_\mathrm{S}\right) l}  \,,
\end{align}
with:
\begin{equation}
\beta= \frac{4\gamma g^2(0)\left(2K_\mathrm{C}-\frac{K_\mathrm{S}}{2}+\frac{1}{2K_\mathrm{S}}\right)}{4K_\mathrm{C}-K_\mathrm{S}+\frac{1}{K_\mathrm{S}}-2}\,.
\end{equation}

If $K_\mathrm{C} > 1- \tilde{K}_\mathrm{S}/4-1/(4\tilde{K}_\mathrm{S})$, then $g$ flows to zero because its exponent is negative, and our assumption on $\tilde{K}_\mathrm{S}$ fails: we may expect that $K_\mathrm{S}$ remains larger then one, driving also $h$ to zero.

If $K_\mathrm{C} < 1- \tilde{K}_\mathrm{S}/4-1/(4\tilde{K}_\mathrm{S})$, the operators $\mathcal{O}$ become relevant. For $\tilde{K}_\mathrm{S} <1$ we must distinguish the following cases:
\begin{itemize}
\item For $K_\mathrm{C} > \frac{1}{2}-\frac{1}{4\tilde{K}_\mathrm{S}}+\frac{\tilde{K}_\mathrm{S}}{4}$, $h$ grows faster than $g^2$ and it is responsible for the opening of a gap which may be estimated by considering $h(l)\approx 1$ (in the energy scale of the bandwidth). This is the case corresponding to the $K=8$ resonance state analyzed in the main text.
 \item For $K_\mathrm{C} < \frac{1}{2}-\frac{1}{4\tilde{K}_\mathrm{S}}+\frac{\tilde{K}_\mathrm{S}}{4}$, $h \propto g^2$ asymptotically: in this case $g$ reaches values of order 1 faster than $h$. This corresponds to a regime where the operators $\mathcal{O_{2,\pm 1}}$ dominate and the system may flow to a different fixed point.
 \end{itemize}

\section{Approximate evaluation of the boundary conditions for the RG flow} \label{sec:initial}

The second order RG equations we found, Eqs. \eqref{reng}, \eqref{renh} and \eqref{renk}, are supposed to qualitatively describe the behavior of the scaling of $g,h$ and $K_\mathrm{S}$ for small values of $\Omega$ and the interactions. Here we provide an approximate evaluation of their ``bare'' initial conditions at $l=0$.

The value of $g(l=0)$ is related to the non-universal coefficients $\alpha_p$ which may be introduced in the sum of the terms \eqref{a} which defines the fermionic operators on the chain. In particular, $g(0) = \Omega \alpha_{p=1} \alpha_{p=3} /(2\pi)$, and, for the sake of simplicity we impose $g(0)=\Omega/(2\pi)$ for the resonant case.

The initial conditions of $h$ and $K_\mathrm{S}$ instead, are strictly related to the repulsive interactions in the system. We consider the following Hubbard and nearest-neighbor interactions:
\begin{equation}
H_{UV}= \sum_r \frac{U}{2}n_r(n_r-1) + V n_r n_{r+1}
\end{equation}
where $n_r= a^\dag_{\Up,r}a_{\Up,r} + a^\dag_{\Dn,r}a_{\Dn,r}$ is the total occupation of the site $r$ of the chain.
We can translate this interaction by considering the first harmonic only in the definition of the fermionic field. We obtain:
\begin{equation}
a^\dag_{\sigma,r}a_{\sigma,r}= \frac{\partial_x \theta_\sigma}{\pi} + \frac{1}{2\pi}\left(ie^{i2k_\mathrm{F} r + 2i\theta_\sigma} -i e^{-i2k_\mathrm{F} r - 2i\theta_\sigma}\right) \,.
\end{equation}
Therefore:
\begin{multline}
U a^\dag_{\Up,r}a_{\Up,r}a^\dag_{\Dn,r}a_{\Dn,r} \to U\frac{\partial_x \theta_\Up  \partial_x \theta_\Dn}{\pi^2} + \frac{U}{2\pi^2}\cos\left[2\left(\theta_\Up - \theta_\Dn\right)\right] - \frac{U}{2\pi^2} \cos \left[2\left(\theta_\Up + \theta_\Dn\right)+4k_\mathrm{F}r\right]  =\\
= \frac{U}{2\pi^2} \left[(\partial_x \theta_\mathrm{C})^2 - (\partial_x \theta_\mathrm{S})^2\right] + \frac{U}{2\pi^2}\cos{2\sqrt{2} \theta_\mathrm{S}} - \frac{U}{2\pi^2}\cos\left(2\sqrt{2} \theta_\mathrm{C} + 4k_\mathrm{F}r\right) \label{Uterm}
\end{multline}
where the umklapp term oscillating with $4k_\mathrm{F} r$ can be neglected for $k_\mathrm{F} \neq \pi/2$, but it can radically change the physics at half filling.
The nearest neighbor interaction reads instead:
\begin{multline}
V n_r n_{r+1} \to \frac{2V}{\pi^2} (\partial_x \theta_\mathrm{C})^2 +\left[ \frac{Ve^{-i2k_\mathrm{F}}}{4\pi^2} \sum_{\sigma,\sigma'}\left(e^{i2\left(\theta_\sigma(r) - \theta_{\sigma'}(r+1)\right)}\right)+{\rm H.c}\right] + \ldots \\
\approx\frac{2V}{\pi^2}\left(1-\cos2k_\mathrm{F}\right) (\partial_x \theta_\mathrm{C})^2 -\frac{V}{\pi^2}\cos2k_\mathrm{F} (\partial_x \theta_\mathrm{S})^2\\
+ \frac{\sqrt{2}V}{\pi^2}\sin 2k_\mathrm{F}\partial_x \theta_\mathrm{C} + \frac{V}{\pi^2}\cos2k_\mathrm{F} \cos2\sqrt{2}\theta_\mathrm{S} + \ldots
\label{Vterm}
\end{multline}
where the dots label terms oscillating as $4k_\mathrm{F}r$ and the term linear in $\partial_x \theta_\mathrm{C}$ can be neglected because it provides an overall energy which depends on the total number of particles only. From the sum of Eqs. \eqref{Uterm} and \eqref{Vterm} we deduce:
\begin{align}
h(l=0)&= \frac{U}{2\pi^2}+ \frac{V}{\pi^2}\cos2k_\mathrm{F} \,,\\
K_\mathrm{S}(l=0)&= \sqrt{\frac{\pi v_F}{\pi v_F-U-2V\cos2k_\mathrm{F}}}>1\,,\\
K_\mathrm{C}(l=0)&=\sqrt{\frac{\pi v_F}{\pi v_F+U+4V\left(1-\cos2k_\mathrm{F}\right)}} <1
\end{align}
with $v_F=2\sin k_\mathrm{F}$ and $k_\mathrm{F}< \pi/4$ for $\phi <\pi$ at the resonance. These results, though, hold only for $U,V \ll \pi v_F$, thus they provide only a qualitative idea of the behavior towards the strongly interacting regime.

\section{The strongly paired phase} \label{app:d}

The interaction term $h\cos\left(2\theta_\Up -2 \theta_\Dn \right) $ has the effect of gapping the spin sector of the chain. Therefore, to analyze its effect it is convenient to separate the gapped spin sector from the gapless charge sector of the Hamiltonian (and we follow the approach in \cite{teokane}). The two sectors are mixed by the original $g$ term of the Hamiltonian.

To study the spin sector,  it is convenient to apply a canonical transformation and redefine:
\begin{equation}
 \varphi_{\mathrm{S}}'=\frac{\varphi_\mathrm{S}}{\sqrt{2}} = \frac{\varphi_\Up - \varphi_\Dn}{2} \,,\quad \theta_\mathrm{S}'=\sqrt{2}\theta_\mathrm{S} = \theta_\Up - \theta_\Dn\,,\quad K'_\mathrm{S}=2K_\mathrm{S}\,.
\end{equation}
In this way the free term of the Hamiltonian of the spin sector remains of the same form,
\begin{equation}
 H_{0,s}=\frac{1}{2\pi}  \int dx \, v_\mathrm{S} K'_\mathrm{S} \left(\partial_x \varphi_\mathrm{S}' \right)^2 + \frac{v_\mathrm{S}}{K'_\mathrm{S}} \left(\partial_x \theta_\mathrm{S}' \right)^2\,,
\end{equation}
and the interaction term \eqref{pairingint} becomes:
\begin{equation}
 h\cos\left( 2\theta_\Up - 2\theta_\Dn \right) = h \cos\left( 2 \theta'_\mathrm{S} \right)\,.
\end{equation}
We can refermionize this sector of the Hamiltonian by defining the Dirac fermions:
\begin{equation}
 \chi^{R/\mathrm{L}} = e^{i\left(\varphi_\mathrm{S}'\pm\theta_\mathrm{S}' \right) } \,,
\end{equation}
such that the Hamiltonian of the spin sector, including the interaction $h$, becomes:
\begin{equation}
 H_\mathrm{S}= \int dx \, iv_\mathrm{S} \left(\chi^{\mathrm{L}\dag} \partial_x \chi^\mathrm{L} -  \chi^{\mathrm{R}\dag} \partial_x \chi^\mathrm{R}\right) + i m_h \left( \chi^{\mathrm{R} \dag} \chi^\mathrm{L} - \chi^{\mathrm{L} \dag} \chi^\mathrm{R} \right) \,,
\end{equation}
with $m_h$ proportional to $h$.

Therefore the spin sector of the system is described by a free and massive Dirac fermion; following the approach in \cite{teokane} we can split the fields $\chi^{\mathrm{R}/\mathrm{L}}$ into pairs of Majorana operators: $\chi^{\mathrm{R}/\mathrm{L}}=( \gamma_1^{\mathrm{R}/\mathrm{L}} -i \gamma_2^{\mathrm{R}/\mathrm{L}}) /2$ such that the Hamiltonian can be expressed as:
\begin{equation}
 H_\mathrm{S}= \int dx \,  \sum_{j=1,2} \left[i\frac{v_\mathrm{S}}{4}\left( \gamma^{\mathrm{L}}_j \partial_x \gamma^\mathrm{L}_j -  \gamma^{\mathrm{R}}_j \partial_x \gamma^\mathrm{R}_j\right)  + i \frac{m_h}{2}  \gamma^{\mathrm{R} }_j \gamma^\mathrm{L}_j  \right]\,.
\end{equation}
These Majorana fields are gapped and their correlation function decays exponentially with the distance like a Bessel function. We can consider an approximation of the kind:
\begin{equation}
 \left\langle \gamma (x) \gamma (y)\right\rangle \approx \rm{sign}(x-y) e^{-m\left|x-y\right|}.
\end{equation}

So far we discussed the gapped spin sector of the model, concerning the gapless charge sector we may redefine a pair of dual fields as:
\begin{equation}
 \varphi_\mathrm{C}'= {\sqrt{4K_\mathrm{C}}}{ \varphi_\mathrm{C}} \,, \quad \theta_\mathrm{C}'= \frac{\theta_\mathrm{C}}{\sqrt{4K_\mathrm{C}}} \,,
\end{equation}
such that the two point correlation function reads:
\begin{equation}
 \left\langle \varphi_\mathrm{C}' (x) \varphi_\mathrm{C}' (y)\right\rangle = - 2\ln \left|x-y\right|\,.
\end{equation}

Based on these assumptions, we can assume that the original electron operators of the model at the resonance are of the form
\begin{equation}
\psi(x)= \gamma(x)e^{i\varphi'_\mathrm{C}(x)}\,,
\end{equation}
where $\gamma$ is a suitable linear combination of the gapped Majorana modes describing the spin sector.
By applying the usual analogy between CFT correlation functions and wavefunctions of the system we may suppose that:
\begin{equation}
 \Psi=\left\langle \psi(x_1)\psi(x_2)\ldots \psi(x_N)\right\rangle \propto {\rm Pf}\left[{\rm{sign}}(x_j-x_k) e^{-m_h\left|x_j-x_k\right| } \right]  \prod_{j<k} \left(x_j-x_k \right)^2\,,
\end{equation}
which constitutes a fermionic state at filling $1/2$.
We can rewrite this expression to account more explicitly for the strong localization of the pairs given by the mass gap $m_h$. We consider the permutations $P$ of the particles and we can approximate the previous expression by:
\begin{multline}
 \Psi \propto \sum_{P} (-1)^P \prod_{j=1}^{N/2} \left[{\rm sign}\left( x_{P(2j)}-x_{P(2j-1)}\right)e^{-m_h\left|x_{P(2j)}-x_{P(2j-1)}\right| }  \right]\\
 \prod_{j<k}^{N/2} \left(\frac{x_{P(2j)}+x_{P(2j-1)}}{2} - \frac{x_{P(2k)}+x_{P(2k-1)}}{2}  \right)^{8}
\end{multline}
where $(-1)^P$ is the parity of the permutation $P$ and we considered only the dependence on the center of mass of each bosonic pair in the Laughlin-Jastrow factor. This reflects the strong exponential localization of the pairs. This wavefunction describes indeed a 1D limit for a Laughlin state at filling $1/8$ of bosonic molecules composed by pairs of fermions \cite{greiter91}, and it is often called the $K=8$ state \cite{wen08}.

 The main observable which detects the appearance of this strongly correlated state is the chiral current. Numerically, we studied it as a function of the magnetic flux $\phi$ close to the $\nu=1/2$ resonance. To give a qualitative description of its behavior we must consider a situation in which we have a small displacement of the flux $\delta \phi = 4k_F - \phi$ around $\phi=4k_F$. In this case it is possible to correct the evaluation of the effective action in Sec. \ref{RG} by considering $\Phi_{2,\pm 1} \to \Phi_{2,\pm 1} +\delta\phi x$. This correction becomes particularly important in the kinetic term in Eqs. (\ref{kin},\ref{kin2}). In particular Eq. \eqref{kin2} results:
\begin{multline} \label{kin3}
S_0 + \int d^2x' {g^2} \alpha^2 D_g dl \left[(\partial_x \Phi_{2,-1} + \delta\phi)^2  + (\partial_x \Phi_{2,+1} + \delta\phi)^2\right]= \\
S_0 + 4\int d^2x' {g^2} \alpha^2 D_g dl \left[(2\partial_x \theta_{\mathrm{C}} - \partial_x \varphi_{\mathrm{S}} + \delta\phi)^2  + (\partial_x \theta_\mathrm{S})^2\right]\,.
\end{multline}
This contribution to the kinetic energy qualitatively justifies a behavior of the expectation value of the chiral current proportional to:
\begin{equation} \label{eq:ohr_current}
\left\langle j_c\right\rangle \propto \left\langle \partial_x \varphi_\mathrm{S}\right\rangle \propto D_g \left(4k_F - \phi \right).
\end{equation}

\section{Correlation functions}
\label{app:correlations}
Here we write the bosonization description of the correlations functions adopted in the text. Let us start from the single-site operators and we approximate them using only the first harmonic $(p=p'=1)$:
\begin{align}
a^\dag_{\Up,r} a_{\Dn,r}  &\propto \; e^{-i\sqrt{2}\left(\varphi_\mathrm{S} - \theta_\mathrm{S}\right) -i\phi r} + e^{-i\sqrt{2}\left(\varphi_\mathrm{S} + \theta_\mathrm{S}\right) -i\phi r} 
+ e^{-i\sqrt{2}\left(\varphi_\mathrm{S} - \theta_\mathrm{C}\right) -i\phi r +i 2k_\mathrm{F}r} + e^{-i\sqrt{2}\left(\varphi_\mathrm{S} + \theta_\mathrm{C}\right) -i\phi r-i 2k_\mathrm{F}r}\,,\\
a^\dag_{\Up,r} a^\dag_{\Dn,r}  &\propto \; e^{-i\sqrt{2}\left(\varphi_\mathrm{C} - \theta_\mathrm{C}\right) +i2k_\mathrm{F}r} + e^{-i\sqrt{2}\left(\varphi_\mathrm{C} + \theta_\mathrm{C}\right) -i2k_\mathrm{F}r} + e^{-i\sqrt{2}\left(\varphi_\mathrm{C} - \theta_\mathrm{S}\right) } + e^{-i\sqrt{2}\left(\varphi_\mathrm{C} + \theta_\mathrm{S}\right) }\,,
\end{align}
where we neglected the Klein factors as well. We approximate the two point correlation functions by considering only the terms which depend on the relative distance $r_2 - r_1$:
\begin{align}
C_{\rm s}(r_2 - r_1) \propto& \left\langle e^{i\sqrt{2}\left(\varphi_\mathrm{S}(r_2) - \varphi_\mathrm{S}(r_1)\right)}\right\rangle\left[e^{i\phi(r_2-r_1)}\left(\sum_{\mu,\nu=\pm 1} \left\langle e^{i\sqrt{2}\left(\mu\theta_\mathrm{S}(r_1) + \nu\theta_\mathrm{S}(r_2)\right)}\right\rangle\right) \right. \nonumber\\
&\left. +  \left\langle e^{i(\phi-2k_\mathrm{F})(r_2-r_1)}e^{i\sqrt{2}\left(\theta_\mathrm{C}(r_1)-\theta_\mathrm{C}(r_2)\right)} + e^{i(\phi+2k_\mathrm{F})(r_2-r_1)}e^{-i\sqrt{2}\left(\theta_\mathrm{C}(r_1)-\theta_\mathrm{C}(r_2)\right)}\right\rangle\right]
\nonumber\\
&- \left\langle  a^\dag_{\Up,r_1} a_{\Dn,r_1}\right\rangle  \left\langle a_{\Up,r_2}a^\dag_{\Dn,r_2}\right\rangle
\end{align}
In the integer resonance $\varphi_\mathrm{S}$ and $\theta_\mathrm{C}$ are pinned to their semiclassical minima and their contribution is erased by the single-site averages, whereas for $\nu=1/2$ the $\varphi_\mathrm{S}$ correlation function is exponentially decaying because of the gap opened by the $h$ term. We get:
\begin{equation}
C_{\rm s}(r_2 - r_1) \propto \left\{
\begin{array}{ll}
e^{i\phi(r_2-r_1)}|r_2-r_1|^{-K_\mathrm{S}} & {\rm for} \quad \nu=1 \\
e^{-m_h|r_2-r_1|}f(r_2-r_1)& {\rm for} \quad \nu=1/2\\
e^{i\phi(r_2-r_1)}\left[|r_2-r_1|^{-K_\mathrm{S}-1/K_\mathrm{S}}\right.\\
\left.\phantom{e^{i\phi(r_2-r_1)}} + \cos\left[2k_\mathrm{F}(r_2-r_1)\right] |r_2-r_1|^{-K_\mathrm{C}-1/K_\mathrm{S}}\right] & {\rm for} \quad \nu\neq 1,1/2
\end{array}
\right.
\label{eq:corr_s}
\end{equation}
with $f$ an algebraically decaying function. Here we considered the most relevant contributions and we neglected higher harmonic terms.
Concerning the pair correlation functions we obtain:
\begin{multline}
C_{\rm p}(r_2 - r_1) \propto \left\langle e^{i\sqrt{2}\left(\varphi_\mathrm{C}(r_2) - \varphi_\mathrm{C}(r_1)\right)}\right\rangle\left[\left\langle e^{i2k_\mathrm{F}(r_2-r_1)}e^{i\sqrt{2}\left(\theta_\mathrm{C}(r_2)-\theta_\mathrm{C}(r_1)\right)} + {\rm H.c.}\vphantom{\sum_{\mu,\nu=\pm 1}} \right\rangle\right.\\\left. + \sum_{\mu,\nu=\pm 1} \left\langle e^{i\sqrt{2}\left(\mu\theta_\mathrm{S}(r_1) + \nu\theta_\mathrm{S}(r_2)\right)}\right\rangle  \right]
\end{multline}
In the integer case the correlations of $\varphi_\mathrm{C}$ decay exponentially because of the gap opened by $\Omega$, therefore we get:
\begin{equation}
C_{\rm p}(r_2 - r_1) \propto \left\{
\begin{array}{ll}
e^{-m_\Omega|r_2-r_1|}g(r_2-r_1) & {\rm for} \quad \nu=1 \\
|r_2-r_1|^{-K_\mathrm{S}-1/K_\mathrm{C}} + \cos\left[2k_\mathrm{F}(r_2-r_1)\right] |r_2-r_1|^{-K_\mathrm{C}-1/K_\mathrm{C}} & {\rm for} \quad \nu\neq 1,1/2\\
 |r_2-r_1|^{-1/K_\mathrm{C}} + \cos\left[2k_\mathrm{F}(r_2-r_1)\right] |r_2-r_1|^{-K_\mathrm{C}-1/K_\mathrm{C}}& {\rm for} \quad \nu=1/2
\end{array}
\right.
\label{eq:corr_p}
\end{equation}
where $g$ is an algebraically decaying function.

\section{Further numerical evidence of the $\nu=1/2$ phase}
\label{sec:further_evidence}

In the main text, we use the chiral current $j_c$ as a quantity to confirm the existence of a gap at $\nu=1/2$. We observe that, in the data we present in Fig. \ref{Fig2}, the downward peak of the current seems to scale to zero with the size of the system. We claim, however, that this is not a signal of the disappearance of the double-cusp pattern of the chiral current in the thermodynamic limit; it is instead a discretization effect due to the limited number of available points within the resonance in our numerical simulations, which depends on the system size, combined with the flux quantization. Most of the observables of the system, including the chiral current, display indeed an oscillating behavior as a function of the flux $\phi$ with period $2\pi/L$. This makes it impossible to compare results of fractional fluxes with results of integer ones, thus limiting the number of points numerically available inside each resonance.

In particular, to compare systems with different $\phi$, we choose to consider always an integer number of fluxes in the full system. This determines that $\phi$ can be varied only by steps of $2\pi/L$, and, for the system sizes available in our numerical simulations, we can obtain only two points in proximity of the double cusp pattern. Therefore depending on how far or close these points are from the true position of the cusps, the numerical data exhibit a larger or smaller discontinuity in $j_c$.
For this reason, any attempt of a finite size extrapolation for the present data would not be rigorous: for different system sizes, the current is measured at different positions in the $1/\nu$ axis. Instead, the sign inversion in proximity of the resonance (but sufficiently far from the cusps) show a much more stable behavior, supporting our argumentation.

To confirm further our claim on the thermodynamic limit of $j_c$, we would then need to perform additional runs with longer chains until we reach one additional point inside the resonant regime of the chiral current, which is beyond our present computational capabilities.
The observables we analyzed in the main text, including central charge, oscillations in the entanglement entropy and pair correlation functions, confirmed however our analytic predictions and can be considered an indirect proof that the scaling of the system to larger sizes $L$ is not detrimental for the measurement of the chiral current.
Here, we present additional data related to the $1/2$ resonance; we show properties of the one-body density matrix (OBDM) and the spectrum of Schmidt-values which clearly indicate a phase transition at $\nu=1/2$ but are not directly linked to our analytic results.

\begin{figure}[h]
	\centering
	\includegraphics[width=.45\textwidth]{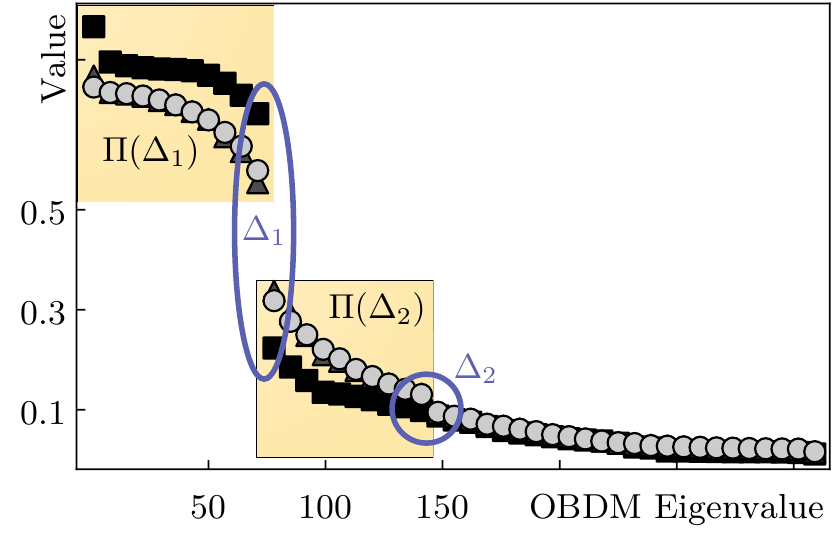}
	\llap{
  \parbox[b]{4cm}{$(a)$\\\rule{0ex}{4cm}
  }}\hfill
	\includegraphics[width=.45\textwidth]{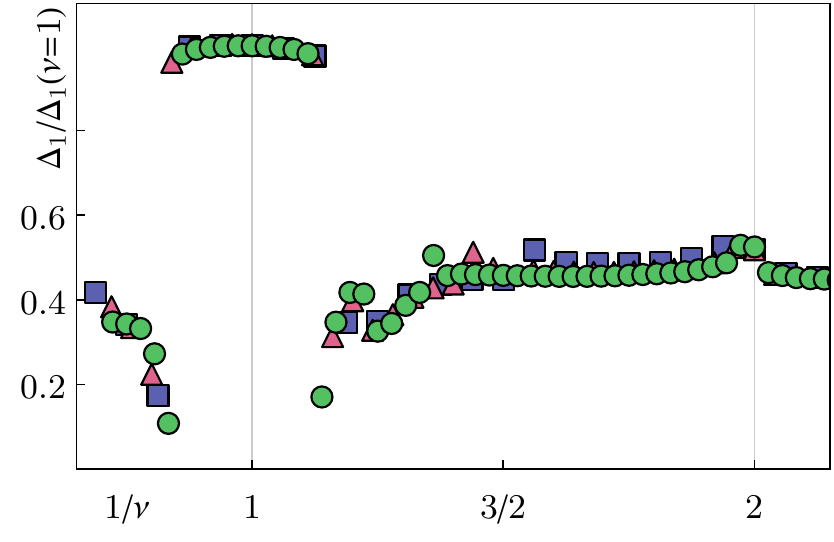}\llap{
  \parbox[b]{4cm}{$(b)$\\\rule{0ex}{4cm}
  }}\\
	\includegraphics[width=.45\textwidth]{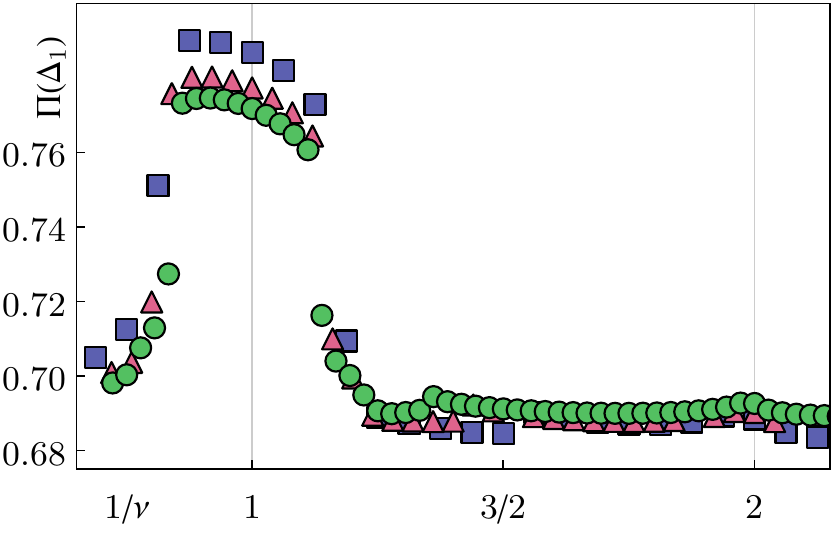}\llap{
  \parbox[b]{4cm}{$(c)$\\\rule{0ex}{4cm}
  }}\hfill
	\includegraphics[width=.45\textwidth]{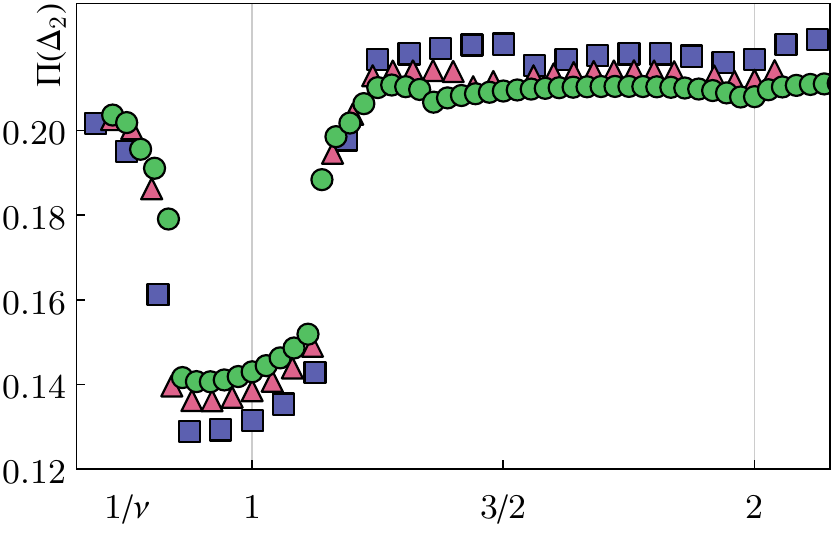}\llap{
  \parbox[b]{5.5cm}{$(d)$\\\rule{0ex}{4cm}
  }}
	\caption{(a) Eigenvalues of the one-body density matrix of the $L/N=156/72$ simulation at $\Omega/t=0.02$ and $U/t=V/t=4$ at $\nu=1$ (black square), $\nu=18/23$ (gray triangle), $\nu=1/2$ (gray disk). (b) Gap $\Delta_1$ for $L/N=72/32$ (blue square) $110/50$ (red triangle) and $156/72$ (green disk), parameters of (a). The $\Delta_1$ gap increases significantly in the integer resonance and at $\nu=1/2$ for all system sizes. The $\Delta_2$ gap, instead, decreases in the integer resonance and at $\nu=1/2$. (c,d) The effects of $\Delta_{1/2}$ manifest in $\Pi(\Delta_{1/2})$ which are means of all eigenvalues in the two yellow colored boxes of (a). The bond dimension used is 400.}
	\label{fig:obdm}
\end{figure}

The off-diagonal elements of the OBDM $\varrho_{r,r'} = \langle a^\dag_{r}a^{\phantom\dag}_{r'}\rangle$ are closely related to the chiral current
\begin{align}
	j_c = -\frac1L\frac\partial{\partial\phi} E_\text{GS} &= \frac {\mathrm it}{2L} \left\langle\sum_r\sigma_z\left(\mathrm e^{\mathrm i\phi/2\sigma_z}a^\dag_r a^{\phantom\dag}_{r+1} - \mathrm e^{-\mathrm i\phi/2\sigma_z}a^{\dag}_{r+1}a^{\phantom\dag}_r\right)\right\rangle
	\label{eq:chiral_current1}
	\\&= -\frac tL \mathrm{Im}\left(\left\langle\sum_r\sigma_z\mathrm e^{\mathrm i\phi\sigma_z/2}a_r^\dag a^{\phantom\dag}_{r+1} \right\rangle\right)
	\label{eq:chiral_current2}
\end{align}
and $\varrho$ manifests two gaps in its eigenvalue spectrum at generic $\nu$ (see Fig.~\ref{fig:obdm} (a)).
We find a large gap $\Delta_1$ and a small gap $\Delta_2$, both sensitive to the integer and the one-half phases.
The gap $\Delta_{1}$ $(\Delta_{2})$ is the difference between the $N$th ($2N$th) and $N+1$th ($2N$+1th) eigenvalues.
$\Delta_1$ at $\nu=1/2$ shows a finite-size scaling which is non-decreasing and stable, supporting our claim for $j_c$ (see Fig.~\ref{fig:obdm} (b)).
\begin{figure}[t]
	\centering
	\includegraphics[width=.6\textwidth]{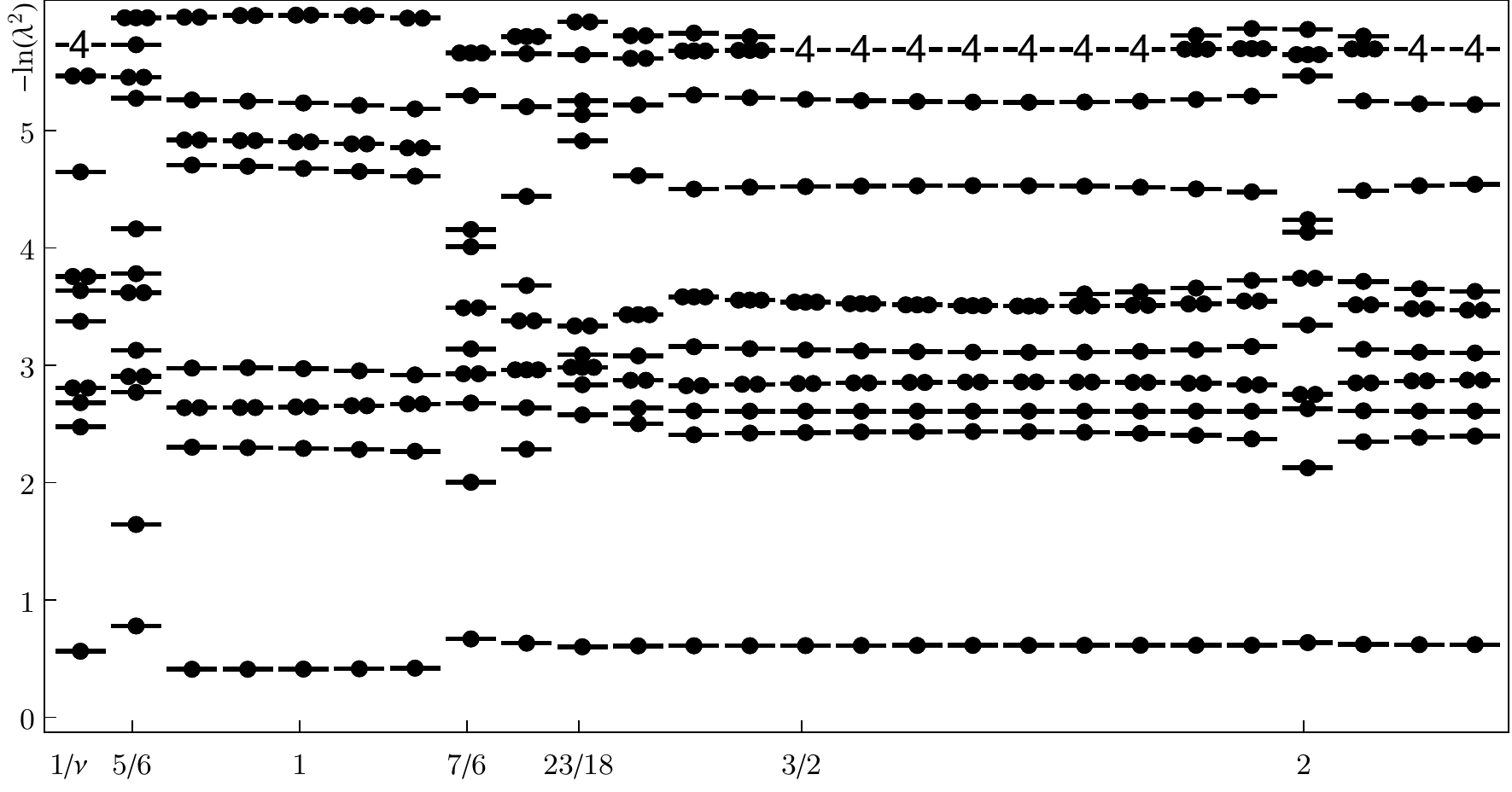}
	\caption{Central-site entanglement spectrum of $L/N=156/72$ at $\Omega/t=0.02$, $U/t=V/t=4$ and bond dimension $400$. Values larger than $6$ are not displayed. In case of $\nu=1/2$, we see a sudden change of the entanglement spectrum, similarly to the two transitions of the $\nu=1$ resonance at $\nu=6/5$ and $\nu=6/7$. The peculiar state at $\nu=18/23$ is subject of a finite-size effect, visible in all observables shown and merging to the integer QH phase at the TL (moving kinks in Fig.~\ref{fig:obdm} (b,c) and (d)).}
	\label{fig:entanglement_spectrum}
\end{figure}\newpage

\section{Error estimates and general remarks about the simulation}
\label{sec:error_discussion}
We use a maximum bond dimension of $M=600$ for all three different system sizes.
As we discuss in the following, this bond dimension is sufficient to capture the physical picture for the present system lengths.
The distance of the two resonant phases as a function of $\phi=N_\phi\frac{2\pi}{L}$ is proportional to the particle filling $N$, leading to $\nu(N)=\frac{N_\phi}{2N}$.
At the same time, the minimum accessible filling is restricted to $\nu_\text{min}=N/L<1/2$ since the energy is symmetric with respect to $\phi\rightarrow2\pi-\phi$.
We decided to simulate densities close to $1/2$ to maximally separate $\nu=1$ from $\nu=1/2$ and at the same time keep track of a few points before the mirroring point $N_\phi=L/2\,$ $(\phi=\pi)$.
Since the accuracy of the approximation depends on the length of the system, we restrict the following investigation to the computationally most expensive setup of $L=156$ sites.\\

We use the chiral current as signature of partial gaps which has been elaborated in detail by many authors as cited in the main text.
Only the upper triangular part of $\varrho_{r,r'}$ has been used to calculate $j_c$ from Eq.~(\ref{eq:chiral_current2}).

\begin{figure}[ht]
	\centering
  	\includegraphics[width=.45\textwidth]{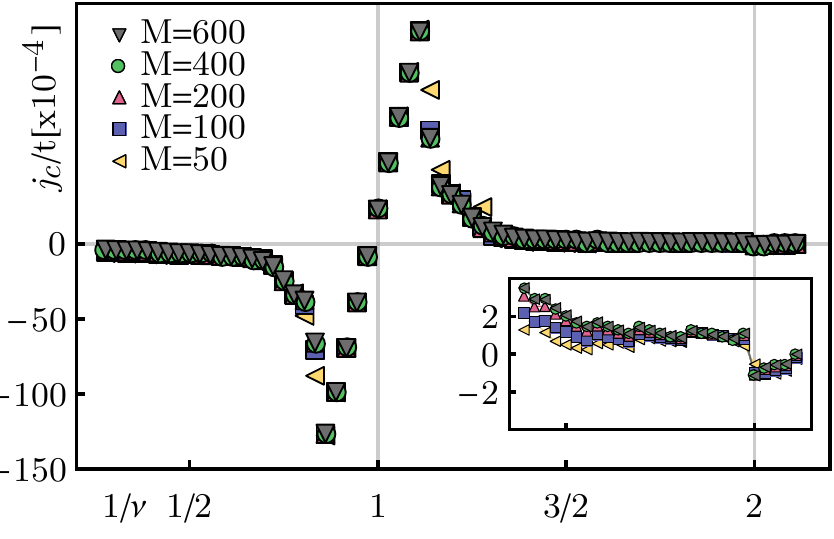}
	\llap{
  \parbox[b]{5cm}{$(a)$\\\rule{0ex}{3.95cm}
  }}\hfill
  	\includegraphics[width=.45\textwidth]{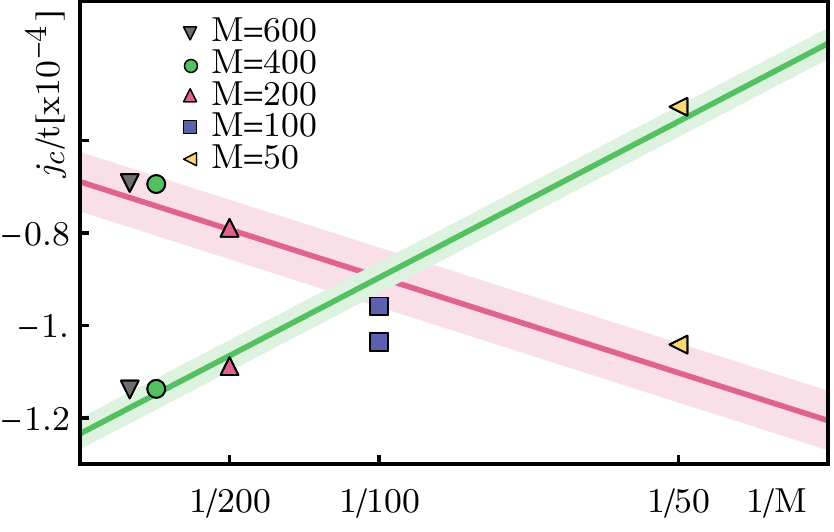}
	\llap{
  \parbox[b]{4cm}{$(b)$\\\rule{0ex}{3.9cm}
  }}\\
  	\includegraphics[width=.45\textwidth]{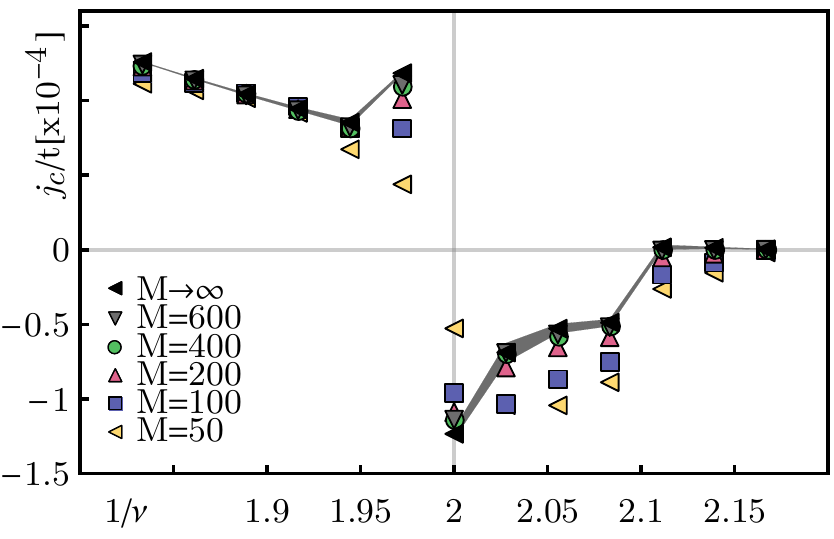}
	\llap{
  \parbox[b]{5cm}{$(c)$\\\rule{0ex}{3.9cm}
  }}\hfill
  	\includegraphics[width=.45\textwidth]{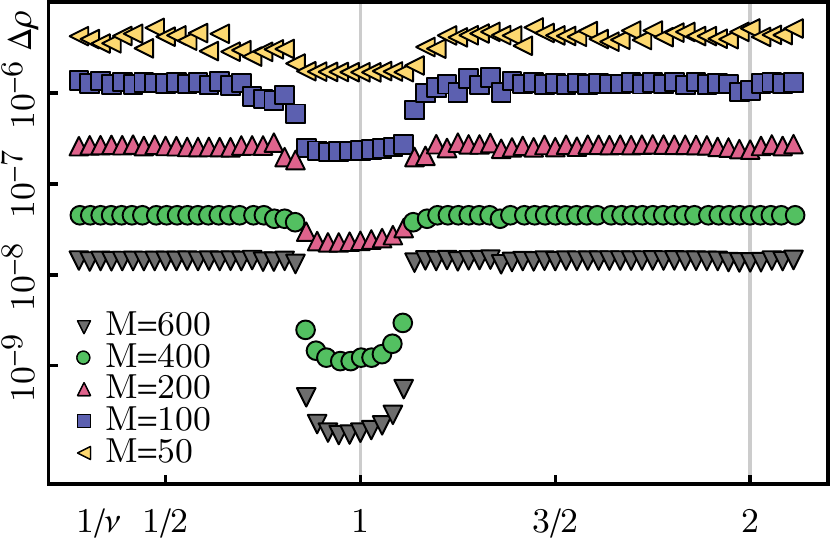}
	\llap{
  \parbox[b]{3cm}{$(d)$\\\rule{0ex}{1.2cm}
  }}
	\caption{
	$L/N=156/72$:
	In (a), we show the finite bond dimension scaling in the full range of $\nu$. To obtain an estimate for the extrapolation $M\rightarrow\infty$ in (c), we show in (b) two results of a linear fit $j_c/t\left(M^{-1}\right) = aM^{-1}+b$ at $\nu=1/2$ (green line) and $\nu=36/73$ (red line). Panel (c) shows the resulting extrapolation in the vicinity of $\nu=1/2$. To notice the error bars of the extrapolation, we use a gray colored area.
	In figure (d) we show the truncated probabilities for different bond dimension. All runs for $U=V=4$ in the main article have a discarded weight $\Delta\rho<10^{-7}$.}
	\label{fig:errors}
\end{figure}

\begin{figure}[ht]
	\centering
  	\includegraphics[width=.32\textwidth]{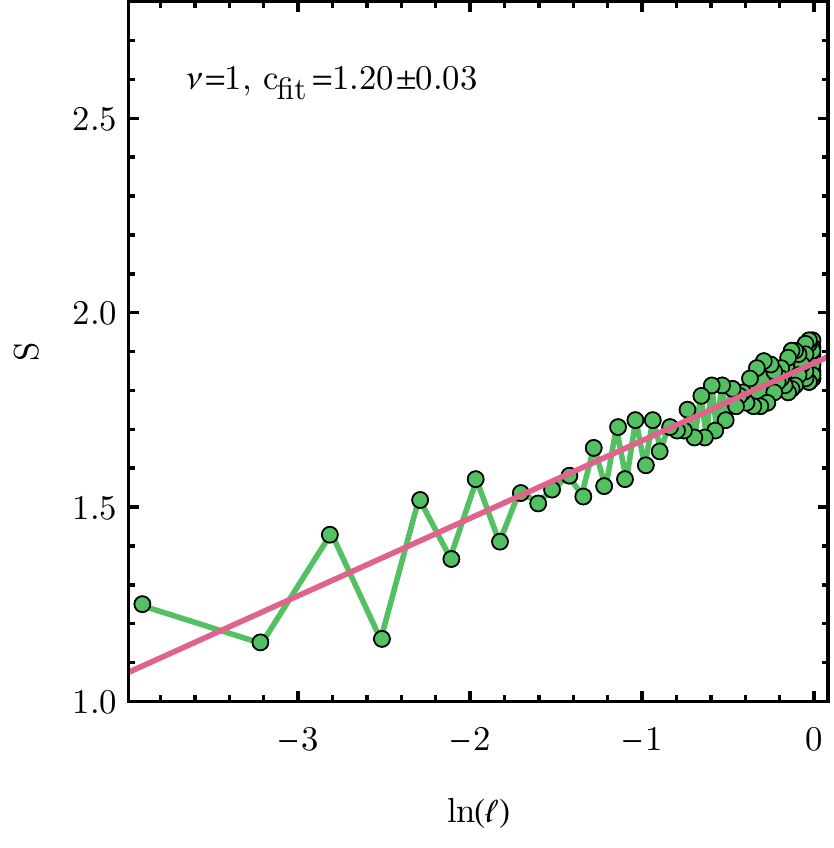}
	\llap{
  \parbox[b]{3cm}{$(a)$\\\rule{0ex}{3.9cm}
  }}\hfill
  	\includegraphics[width=.32\textwidth]{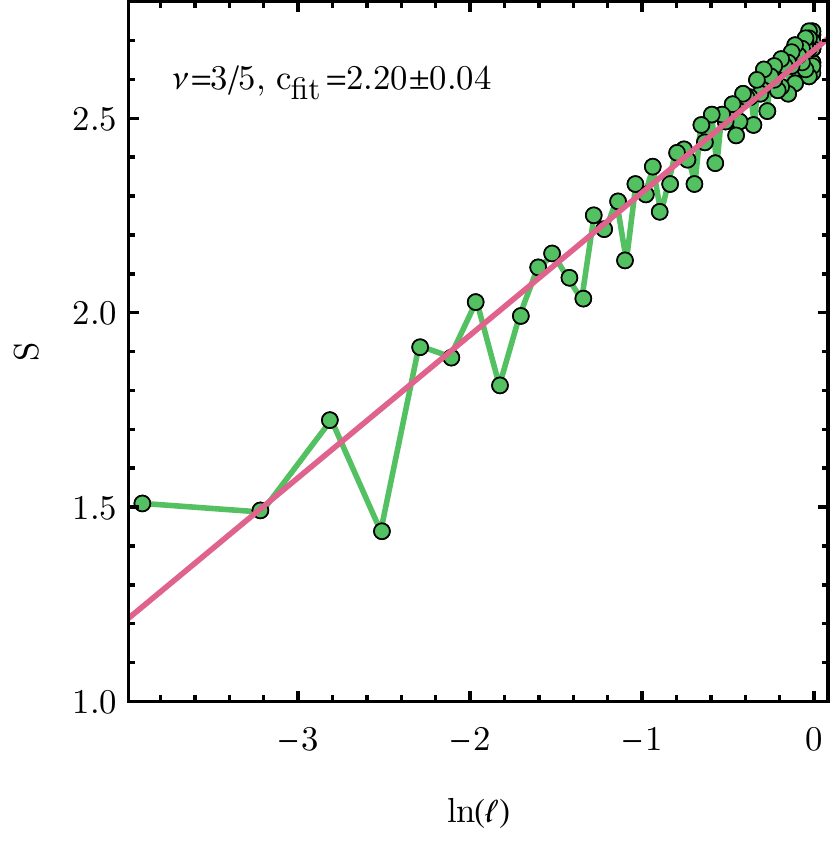}
	\llap{
  \parbox[b]{3cm}{$(b)$\\\rule{0ex}{3.9cm}
  }}
  	\includegraphics[width=.32\textwidth]{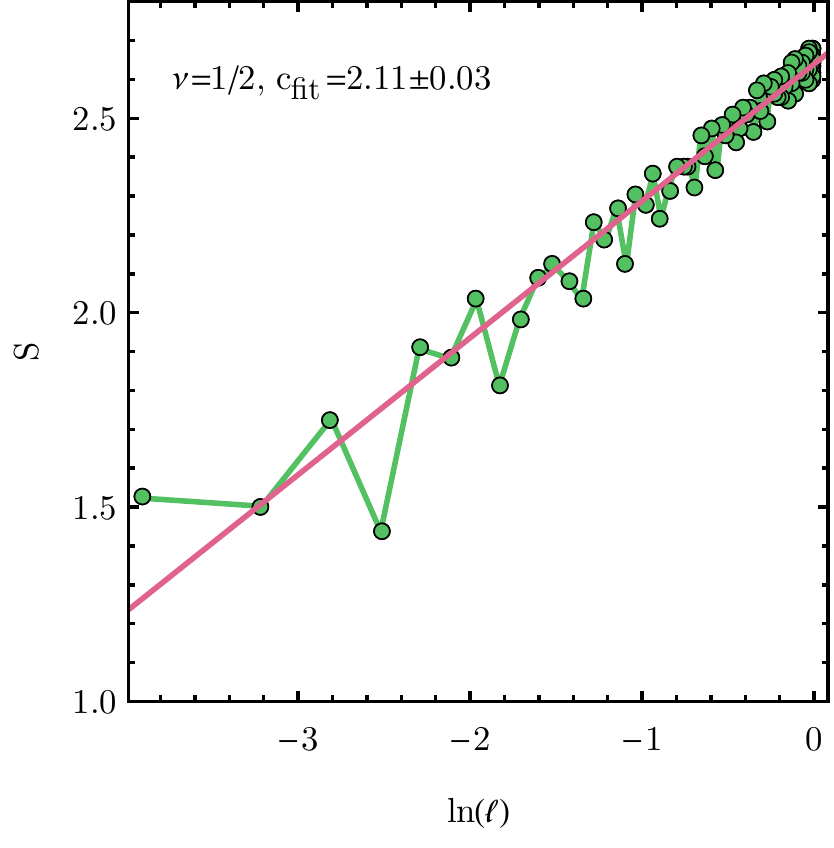}
	\llap{
  \parbox[b]{3cm}{$(c)$\\\rule{0ex}{3.9cm}
  }}\\
  \includegraphics[width=.32\textwidth]{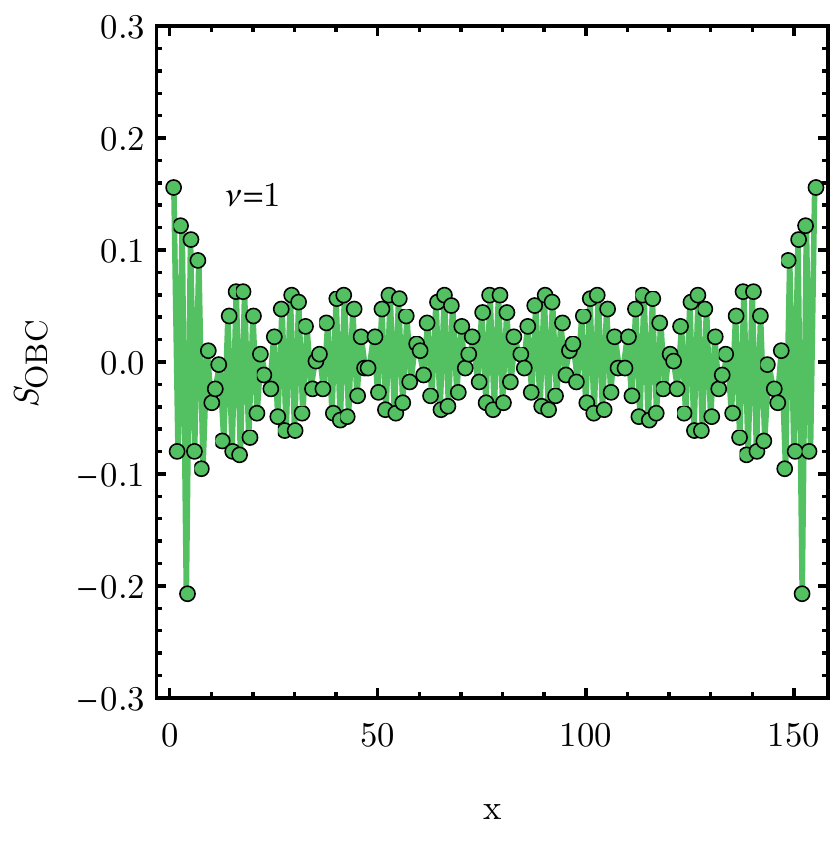}
	\llap{
  \parbox[b]{3cm}{$(d)$\\\rule{0ex}{3.9cm}
  }}\hfill
  	\includegraphics[width=.32\textwidth]{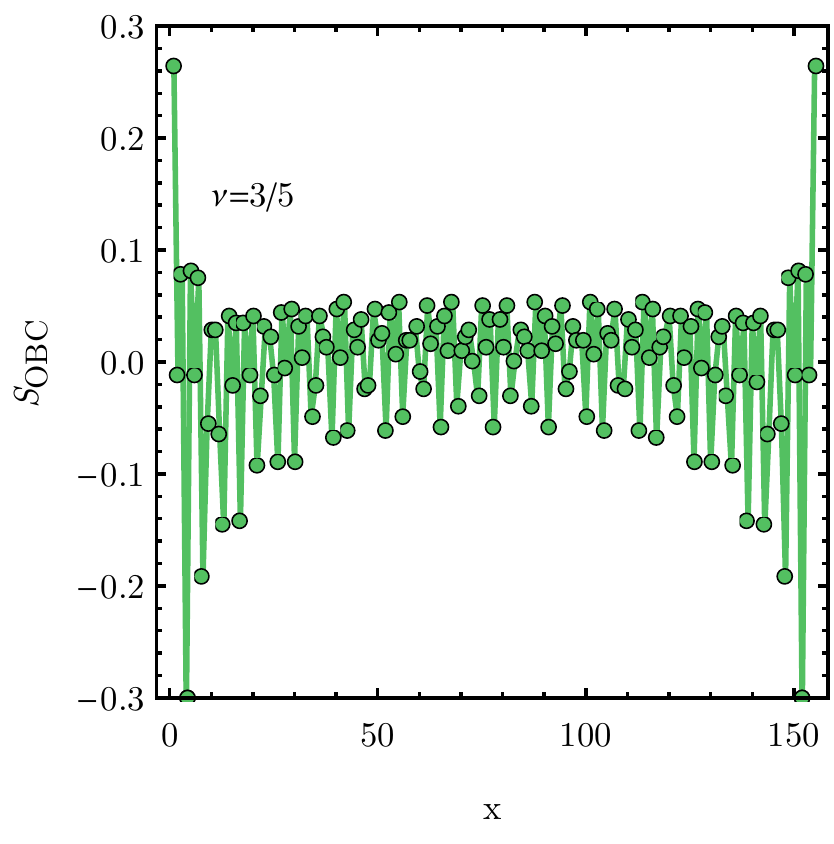}
	\llap{
  \parbox[b]{3cm}{$(e)$\\\rule{0ex}{3.9cm}
  }}
  	\includegraphics[width=.32\textwidth]{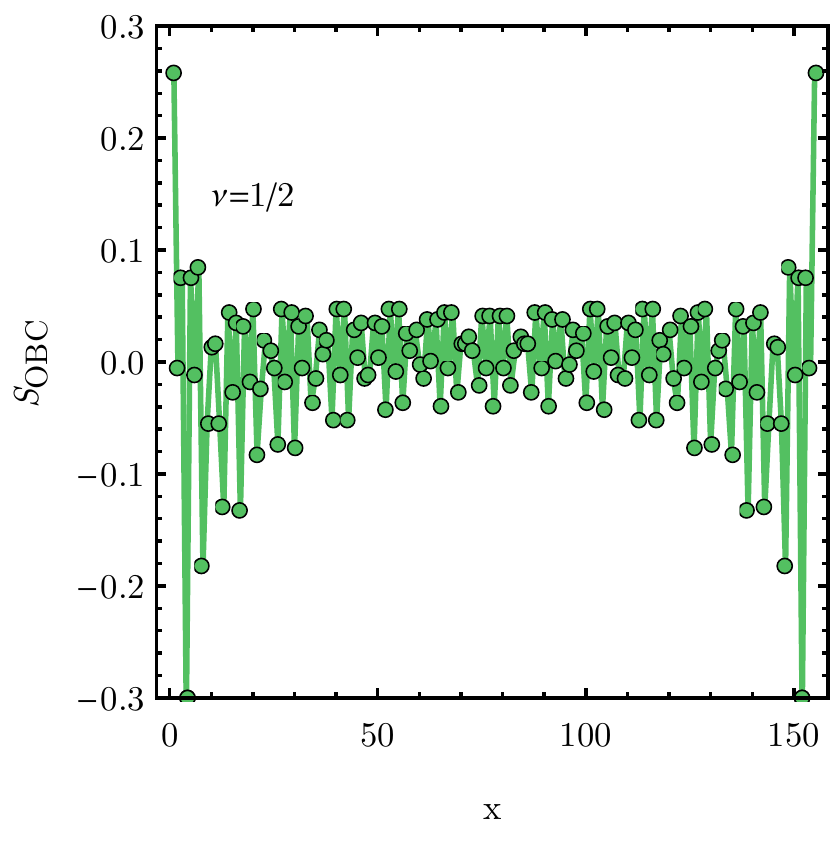}
	\llap{
  \parbox[b]{3cm}{$(f)$\\\rule{0ex}{3.9cm}
  }}\\
  \includegraphics[width=.32\textwidth]{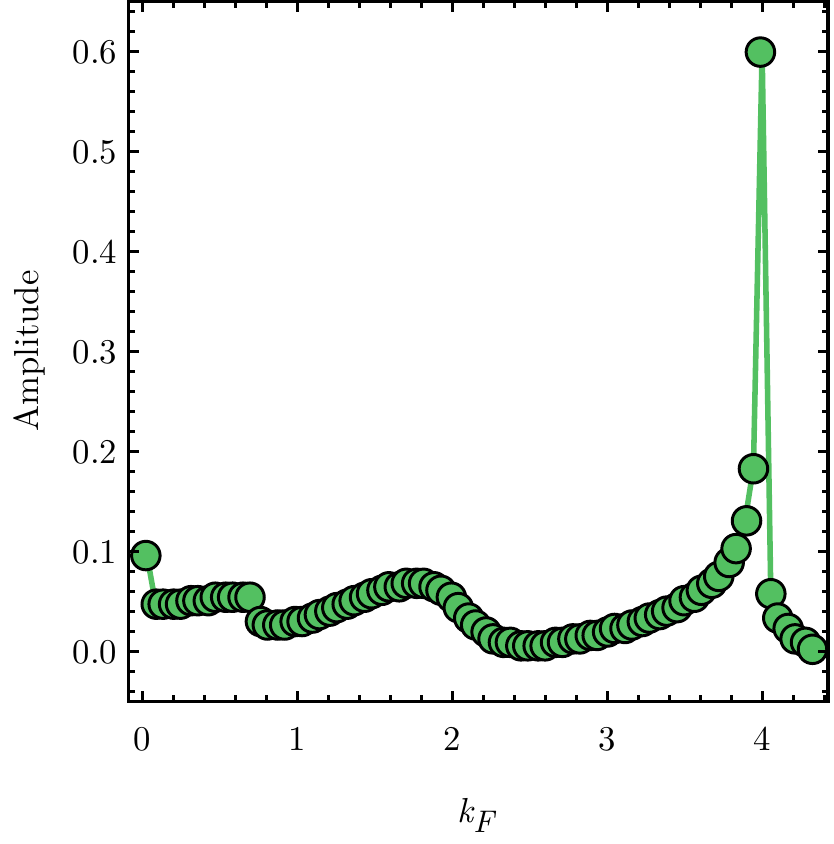}
	\llap{
  \parbox[b]{3cm}{$(g)$\\\rule{0ex}{3.9cm}
  }}\hfill
  	\includegraphics[width=.32\textwidth]{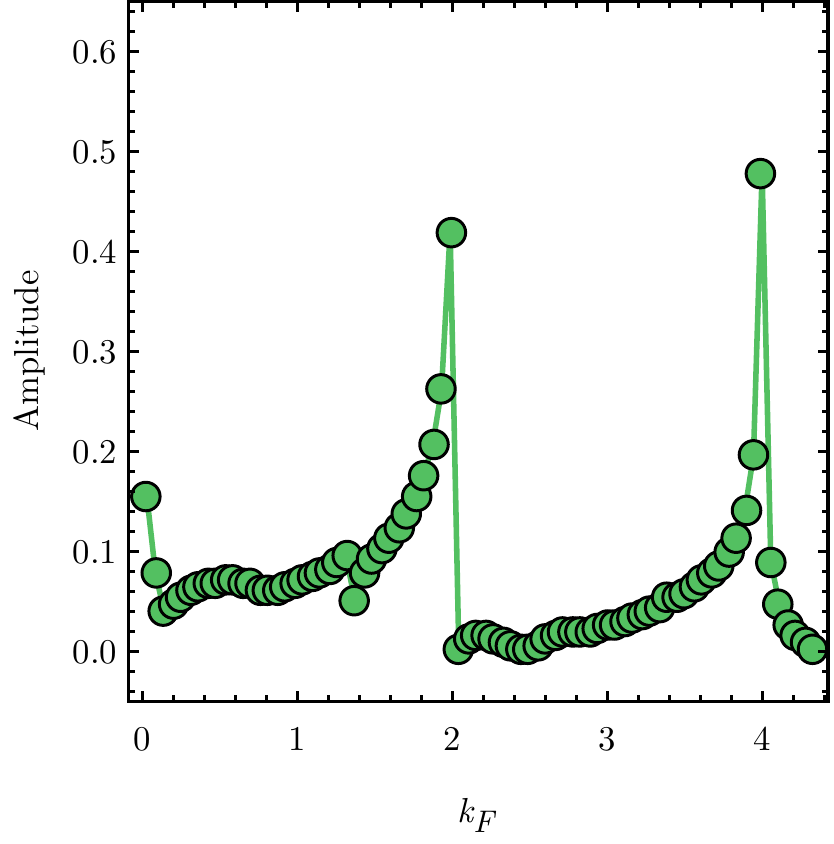}
	\llap{
  \parbox[b]{3cm}{$(h)$\\\rule{0ex}{3.9cm}
  }}
  	\includegraphics[width=.32\textwidth]{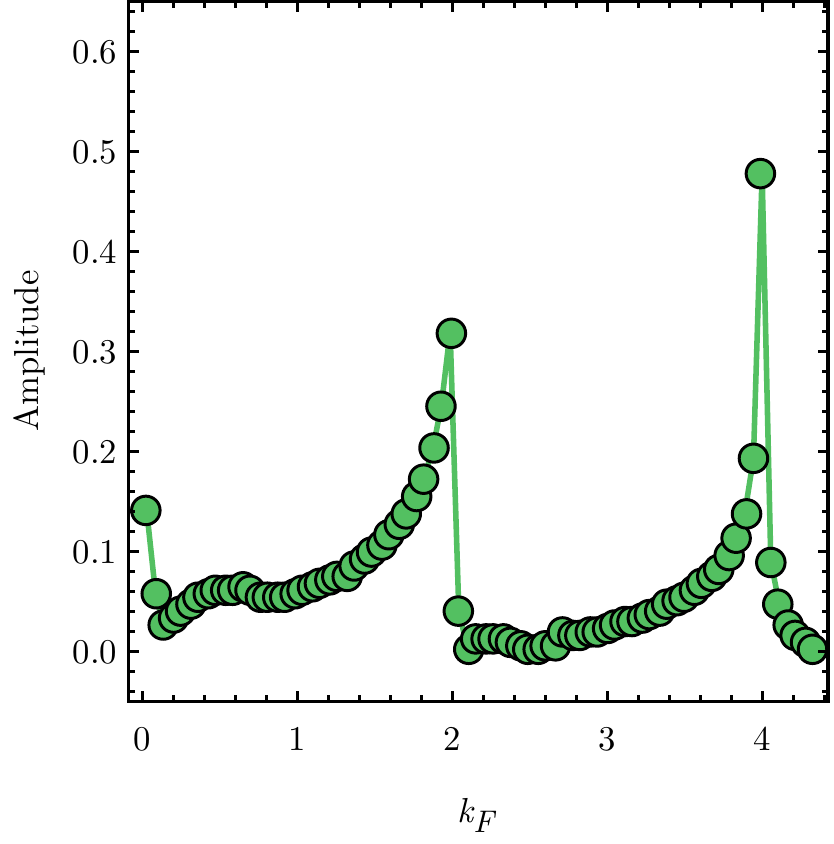}
	\llap{
  \parbox[b]{3cm}{$(i)$\\\rule{0ex}{3.9cm}
  }}
	\caption{
	Entanglement entropy versus chord distance $\ell$ for $L/N=156/72$ for different filling factors $\nu$. (a)-(c): We calculated the scaling of the entanglement entropy by fitting the full data set. (d)-(f): to obtain the open boundary corrections we substracted the central charge scaling. (g)-(i): the frequency spectrum of (d)-(f) reveals two dominant contributions at $2$ and $4k_F$. The contribution at $2k_F$ is strongly suppressed at the resonance, consistent with the emergence of a gap in the spin sector.
	}
	\label{fig:central_charge_fit}
\end{figure}
To estimate the approximation error, it is most common to extrapolate the bond dimension $M$ which restrics the number of states being kept in the MPS.
It is equivalently possible to extrapolate the truncation error, which quantifies the weight probability of the wavefunction being truncated during the optimization process.
For completeness, we plot the truncated probabilities of all bond dimensions in Fig.~\ref{fig:errors}~(d).
As shown in Fig.~\ref{fig:errors}~(a) we observe a saturated scaling of the chiral current as a function of $M$ inside the linear $\nu=1$ region, whereas in case of $\nu=1/2$ the peaks {\em increase} with the bond dimension.
We show a paradigmatic extrapolation at $\nu=1/2$ and $\nu=36/73$ in Fig.~\ref{fig:errors}~(b) and zoom into the fractional resonance in (c).\\

The entanglement entropy has been analyzed according to the thermodynamic Calabrese-Cardy scaling as cited in the main text.
When we eliminate this scaling from the data, we obtain corrections caused by the boundary conditions. The frequency spectrum of these oscillations yields two dominant contributions at $2k_F$ and $4k_F$ (see Fig.~\ref{fig:central_charge_fit} for three examples).


\begin{thebibliography}{44}
\bibitem{footnote1}
An analogous situation is obtained through particle-hole transformation in the large density case.
\end{thebibliography}
\end{document}